\documentclass[11pt,a4paper]{article}
\usepackage{epsfig}
\usepackage[T1]{fontenc}    % Accents cods dans la fonte.
\usepackage{graphics}
\usepackage{graphicx}
\usepackage{pstricks,pst-coil,pst-fill,pst-plot}
\usepackage[fleqn]{amsmath}    % Les symboles les plus frquents.
\usepackage{amssymb}    % Des symboles.
\usepackage{amsfonts}   % Des fontes, eg pour \mathbb.
\usepackage{verbatim}   % Pour les codes sources en informatique.
\usepackage{mathrsfs}   % Des lettres majuscules cursives (\mathscr).
\usepackage{dsfont}
\usepackage{euscript}
\usepackage{yfonts}
\usepackage{txfonts}
\usepackage{marvosym}
\usepackage{vmargin}        % Rgler la taille de la feuille.

\setmarginsrb{1.8cm}{2cm}{1.8cm}{2cm}{1cm}{1cm}{1cm}{1.6cm}
 \makeatletter
 \@addtoreset{equation}{section}
 \makeatother

\let\tend=\rightarrow

\providecommand{\bysame}{\leavevmode\hbox to3em{\hrulefill}\thinspace}

\newtheorem{theorem}{Theorem}[section]
\newtheorem{prop}{Proposition}[section]

\newtheorem{lemme}{Lemma}[section]

\def\Proof{\medskip\noindent {\it Proof --- \ }}

\def\qed{\hfill\rule{2mm}{2mm}}

\newcommand\beq{\begin{equation}}
\newcommand\enq{\end{equation}}
\newcommand\bem{\begin{multline}}
\newcommand\enm{\end{multline}}

\def\beqa{\begin{eqnarray}}
\def\eeqa{\end{eqnarray}}
\def\ba{\begin{array}}
\def\ea{\end{array}}

\def\det{\operatorname{det}}

\newcommand{\f}[2]{{\ensuremath{%
    \mathchoice%
    {\dfrac{#1}{#2}}
    {\dfrac{#1}{#2}}
    {\frac{#1}{#2}}
    {\frac{#1}{#2}}
}}}
\newcommand{\tf}[2]{\ensuremath{#1/#2}}
\newcommand{\pa}[1]{\ensuremath{\left(#1\right)}}
\newcommand{\paa}[1]{\ensuremath{\left\{#1\right\}}}
\newcommand{\pac}[1]{\ensuremath{\left[#1\right]}}
\newcommand{\paf}[2]{\ensuremath{\left(\f{#1}{#2}\right)}}
\newcommand{\pab}[2]{\ensuremath{\pa{\ba{c} #1 \\ #2 \ea }}}
\newcommand{\pabb}[3]{\ensuremath{\pa{ #1 \left| \ba{c} #2 \\ #3 \ea \right .}}  }
\def\a{\alpha}
\def\al{\aleph}
\def\be{\beta}
\def\ga{\gamma}
\def\Ga{\Gamma}
\def\de{\delta}

\def\De{\Delta}
\def\eps{\epsilon}

\def\la{\lambda}
\def\La{\Lambda}

\def\sg{\sigma}

\def\th{\theta}

\def\Om{\Omega}
\def\om{\omega}
\def\vp{\varphi}

\newcommand{\mc}[1]{\ensuremath{\mathcal{#1}}}
\newcommand{\mf}[1]{\ensuremath{\mathfrak{#1}}}
\newcommand{\msc}[1]{\ensuremath{\mathscr{#1}}}

\newcommand{\bs}[1]{\ensuremath{\boldsymbol{#1}}}
\newcommand{\ov}[1]{\ensuremath{\overline{#1}}}
\newcommand{\wt}[1]{\ensuremath{\widetilde{#1}}}
\newcommand{\wh}[1]{\ensuremath{\widehat{#1}}}

\newcommand{\Int}[2]{\ensuremath{\int\limits_{#1}^{#2}}}

\newcommand{\sul}[2]{\ensuremath{\sum\limits_{#1}^{#2}}}
\newcommand{\pl}[2]{\ensuremath{\prod\limits_{#1}^{#2}}}

\newcommand{\R}{\ensuremath{\mathbb{R}}}
\newcommand{\Cx}{\ensuremath{\mathbb{C}}}

\newcommand{\Dp}[1]{\ensuremath{\partial_{#1}}}
\newcommand{\limit}[2]{\ensuremath{\underset{#1 \tend #2}{\longrightarrow} }}
\newcommand{\ex}[1]{\ensuremath{\e{e}^{#1}}}

\newcommand{\bra}[1]{\left\langle \,#1\,\right|}
\newcommand{\ket}[1]{\left|\,#1\, \right\rangle}

\newcommand{\braket}[2]{\ensuremath{ \Big< #1 \big| \,  #2 \Big>  }}
\newcommand{\ddet}[2]{\ensuremath{\det_{#1}\pac{#2}}}

\newcommand{\abs}[1]{\ensuremath{\left| #1 \right|}}

\newcommand{\norm}[1]{\ensuremath{\left\|#1\right\|}}

\newcommand{\dd}{\mathrm{d}}
\newcommand{\e}[1]{\ensuremath{\mathrm{#1}}}

\newcommand{\intff}[2]{\ensuremath{\left [ \, #1 \,; #2 \, \right ] }}

\newcommand{\intof}[2]{\ensuremath{\left ] \, #1 \,; #2 \, \right ] }}

\newcommand{\intn}[2]{\ensuremath{[\![ \, #1 \,;\, #2 \,]\!]}}

\begin{document}

\begin{flushright}
DESY 11-072
\end{flushright}
\par \vskip .1in \noindent

\vspace{14pt}

\begin{center}
\begin{LARGE}
{\bf On Form Factors of the conjugated field in the non-linear Schr\"{o}dinger model.}
\end{LARGE}

\vspace{30pt}

\begin{large}

{\bf K.~K.~Kozlowski}\footnote[1]{DESY, Hamburg, Deutschland,
 karol.kajetan.kozlowski@desy.de},~~
\par

\end{large}

\vspace{40pt}

\centerline{\bf Abstract} \vspace{1cm}
\parbox{12cm}{\small Izergin-Korepin's lattice discretization of the non-linear Schr\"{o}dinger model
along with Oota's inverse problem provides one with determinant representations for the form factors of 
the lattice discretized conjugated field operator.  We prove that these form factors converge, in the zero lattice spacing limit, 
to those of the conjugated field operator
in the continuous model. We also compute the large-volume asymptotic behavior of such form factors in the continuous model. 
These are in particular characterized by Fredholm determinants of operators acting on closed contours. 
We provide a way of defining these Fredholm determinants in the case of generic paramateres. }

\end{center}

\vspace{40pt}

\section*{Introduction\label{INT}}

Finite volume lattice discretizations provide a natural way of circumventing problems related with the 
ultraviolet and infrared divergencies of quantum field theories in infinite volume. As such, they offer a possibility
of a rigorous analysis of the spectrum and correlation functions, the strategy being first to obtain expressions for the lattice discretized finite-volume model and then take appropriate 
limits so as to reach the results relative to the continuous models of quantum field theory in infinite volume.
Clearly, in general, carrying out such a program is hopeless
in as much as finite-volume lattice discretizations introduce tremendous complication of the model. Yet, 
in the case of integrable quantum field theories in (1+1) dimensions it has been shown that, for a wide variety of models, 
there do indeed exist finite volume lattice discretizations preserving the integrable structure of the model 
\cite{BytskoTeschnerSinhGordonFunctionalBA,IzerginKorepinLatticeVersionsofQFTModelsABANLSEandSineGordon,RidoutTeschnerIntegrabilityFamilySigmaModels}.  
The latter can be solved either by means of the algebraic Bethe Ansatz  
\cite{DerkachovKorchemskyManashovXXXSoVandQop,IzerginKorepinLatticeVersionsofQFTModelsABANLSEandSineGordon} or 
through the quantum separation of variables 
\cite{BytskoTeschnerSinhGordonFunctionalBA,DerkachovKorchemskyManashovXXXSoVandQop,NiccoliTeschnerSineGordonRevisited}.  
Such methods lead eventually to the caracterization of the spectrum be means of non-linear integral equations 
\cite{DestriDeVegaAsymptoticAnalysisCountingFunctionAndFiniteSizeCorrectionsinTBAFiniteMagField,TeschnerSpectrumSinhGFiniteVolume}.
It is then possible to take the continuous (infinite number of sites) limit  on the level of such non-linear integral equations. 
This gives access to the spectrum of the associated quantum field theory in finite volume. In such a way, it was
shown for several models 
\cite{DestriDeVegaAsymptoticAnalysisCountingFunctionAndFiniteSizeCorrectionsinTBAFiniteMagField,NiccoliSineGordonSoVNLIELimitBootstrap}
that the infinite volume limit of such a description reproduces the predictions \cite{ZalZalBrosFactorizedSMatricesIn(1+1)QFT} 
for the S matrix and the spectrum 
that were building on the factorizable scattering theory in infinite volume.

The purpose of this paper is to push the study of continuous limits of integrable lattice regularizations of quantum
field theories a step further, this time in respect to the correlation functions. We will focus on the simplest 
possible example, the non-linear Schr\"{o}dinger model (NLSM). Starting from its lattice
discretization introduced by Izering and Korepin \cite{IzerginKorepinLatticeVersionsofQFTModelsABANLSEandSineGordon}, 
we recall the inverse problem of Oota \cite{OotaInverseProblemForFieldTheoriesIntegrability}
and Slavnov's scalar product formula \cite{SlavnovScalarProductsXXZ} so as to provide determinant 
representations for the lattice approximation of the conjugated field operator.  
By generalizing and simpifying the approach of \cite{DorlasOrthogonalityAndCompletenessNLSE}, we show that these form factors, 
along with the generic scalar products and norms, 
converge, when the lattice spacing goes to zero, to the associated quantities arizing in the continuous model in finite volume $L$.  
This constitutes the main result of the paper. Our approach can be applied to many other correlators in this model. 
In particular, it provides the missing steps in the derivation of the previously obtained determinant 
representations for the field, conjugate field and current operators 
\cite{KorepinSlavnovFormFactorsNLSEasDeterminants,OotaInverseProblemForFieldTheoriesIntegrability}
in the continuous model. Finally, building on the techniques introduced in \cite{SlavnovFormFactorsNLSE}
and further developed in 
\cite{KozKitMailSlaTerThermoLimPartHoleFormFactorsForXXZ,KozKitMailSlaTerEffectiveFormFactorsForXXZ}  
we compute the large volume $L$ behavior of the properly normalized determinant representation for the 
conjugated field's form factors. These results are alternative to those obtained in \cite{CauxGlazmanImambekovShashiAsymptoticsStaticDynamicTwoPtFct1DBoseGas} in as much as 
we start from different determinant representations. The large volume asymptotics we obtain are used in \cite{KozReducedDensityMatrixAsymptNLSE}
to derive the long-time and large-distance asymptotic behavior of the so-called one particle
reduced density matrix in the NLSM. We also would like to recall that such large $L$ asymptotics of properly 
normalized form factors involve Fredholm determinants of integral operators acting on a closed contour. 
These determinants may fail to be well-defined in the case of arbitrary excited state. In the core of this
paper we provide a way to circumvent such difficulties.

This paper is organized as follows. 
In section \ref{Section NLSE discret}, we introduce the lattice discretisation of the NLSM and recall several known facts 
about the model. In section \ref{Section FF resultat principal}, we present the main result of the paper:
the convergence (in the zero lattice spacing limit) of the form factors for the lattice discretisation of the model to those of 
the continuous model. We also provide determinant representations for these form factors in the continuum. 
In section \ref{Section Thermo limit FF}, we provide formulae for the large volume limit for these form factors in the 
so-called $n$-particle/hole sector. In addition we proove a theorem providing some clarification in respect to 
the definition of the Fredholm determinants occuring in these expressions. 
The proof of theorem \ref{Theorem cvgce lattice discreization}, which is slightly technical, is gathered in appendix
\ref{Appendix Proof continuous limit}.

\section{The lattice discretization of the model}

\label{Section NLSE discret}

\subsection{The Lax matrix}

The Lax matrix proposed by Izergin and Korepin \cite{IzerginKorepinLatticeVersionsofQFTModelsABANLSEandSineGordon} for the lattice non-linear 
Schr\"{o}dinger model reads
\beq
L_{0 n}\pa{\la}= \pa{ \ba{cc}     -i\f{\la}{2}\De + Z_n +c \tf{\chi_n^{*}\chi_n}{2} &  -i\sqrt{c} \chi_n^{*} \rho_{Z_n}  \\
                           i\sqrt{c} \rho_{Z_n} \chi_n  & i\f{\la}{2}\De + Z_n +c \tf{\chi_n^{*}\chi_n}{2}     \ea} \; , \qquad \e{where}
\quad Z_{n}=1+ \pa{-1}^{n} \tf{c\De}{4} \;.                           
\label{ecriture Matrice de Lax}
\enq
It is represented as a $2\times 2$ matrix on the auxiliary space $V_0\simeq \Cx^2 $ whose entries are operators acting on some dense subspace 
of $\mc{H}_n \simeq L^{2}\pa{\R}$. The operators $\chi_n$, $\chi_m^{*}$ are canonical Bose fields with commutation relations 
$\pac{\chi_n,\chi_m^{*}}= \De \de_{n,m}$. In particular, $\chi_n^{*}$ is the adjoint of $\chi_n$ and $\rho_{Z_n}=\sqrt{Z_n+c\tf{\chi^{*}_n\chi_n}{4}}$. 
The parameter $\De$ plays the role of the lattice spacing.

The index $n$ labels the copy of
the quantum space $\mc{H}_n$ where the canonical fields $\chi_n, \, \chi_n^*$ act non-trivially.
It is readily checked that the various fields entering in the definition of the Lax matrix satisfy to the additional relations
\beq
\chi_n \, \rho_{Z_n-\f{\De c}{4}} =  \rho_{Z_n} \,   \chi_n   \quad \qquad \e{and} \quad \qquad  
\rho_{Z_n-\f{\De c}{4}} \,  \chi^*_n =   \chi^*_n \,  \rho_{Z_n}    \;.
\label{definition relation commutation algebrique rhoZ}
\enq
%
%
%
%Note that relation \eqref{definition relation commutation algebrique rhoZ},
%together whith the value of the square of $\rho_{Z_n}$ and the canonical commutation relations can be taken as a closed system
%of defining relations for the algebra generated by $\chi_n, \chi_n^{*}$ and $\rho_{\mf{z}_n}$, with 
%$\mf{z}_n \in Z_n+\tf{\De c \mathbb{Z}}{4}$.

The Lax matrix \eqref{ecriture Matrice de Lax} satisfies the Yang-Baxter equation
\beq
R_{00^{\prime}}\pa{\la-\mu} L_{0n}\pa{\la} L_{0^{\prime}n}\pa{\mu} 
		=   L_{0^{\prime}n}\pa{\mu} L_{0n}\pa{\la}    R_{00^{\prime}}\pa{\la-\mu} \; ,
\label{ecriture YBE}		
\enq
driven by the rational R-matrix $R_{00^{\prime}}\pa{\la} = \la -ic \mc{P}_{00^{\prime}}$, with $\mc{P}_{00^{\prime}}$
being the permutation operator in $V_0\otimes V_{0^{\prime}}$. 
%
%
%
%\beq
%
%R\pa{\la}= \pa{\ba{cccc} \la-ic  & 0 &0 &0 \\  0& \la & -ic & 0 \\ 0 & -ic & \la & 0 \\ 0 & 0 & 0 & \la-ic  \ea} \;.
%
%\enq
%
%
%
The matrix $R\pa{\la}$ becomes a one-dimensional projector at $\la=ic$. As a consequence,  the Lax matrix $L_{0n}\pa{\la}$ satisfies
the quantum determinant relation
\beq
L_{0n}\pa{\la}\sg_{0}^{y} L_{0n}^{t_0}\pa{\la+ic} \sg_{0}^{y} = \f{\De^2}{4} \pa{\la-\f{2iZ_n}{\De}+ic}\pa{\la+\f{2iZ_n}{\De}} =
\f{\De^2}{4} \pa{\la-\nu_n}\pa{\la-\ov{\nu}_n+ic} \quad \e{with} \quad \nu_n=-\f{2iZ_n}{\De} \; .
\enq
Above and in the following, $\ov{z}$ stands for the complex conjugate of $z$. 

It was observed by Izergin and Korepin \cite{IzerginKorepinLatticeVersionsofQFTModelsABANLSEandSineGordon} 
that the zeroes of the quantum determinant define the values of the spectral parameter where the Lax matrix 
has rank one. Namely, the Lax matrix \eqref{ecriture Matrice de Lax}  becomes a direct projector at the points $\nu_n, \ov{\nu}_n-ic$:
\beq
\pac{ L_{0n}\pa{\nu_n} }_{ab}=\a_a^{\pa{+}}\!\pa{n} \, \be_b^{\pa{+}}\!\pa{n} \; , \quad \e{with} \qquad  
\a^{\pa{+}}\!\pa{n}=\pa{\ba{c} \sqrt{c} \, \chi_n^{*} \\ 2i\rho_{Z_n} \ea} \; , \quad
\be^{\pa{+}}\!\pa{n}=\f{1}{2}\pa{\ba{c} \sqrt{c} \, \chi_n \\ -2i\rho_{Z_n} \ea} \;, 
\enq
\beq
\pac{ L_{0n}\pa{\ov{\nu}_n-ic} }_{ab} = \a_a^{\pa{-}}\!\pa{n} \, \be_b^{\pa{-}}\!\pa{n}  \; , \quad \e{with} \qquad  
\a^{\pa{-}}\!\pa{n} = \pa{\ba{c} -2i\rho_{Z_n-\tf{\De c}{4}} \\ \sqrt{c} \, \chi_n \ea} \; , \quad
\be^{\pa{-}}\!\pa{n} = \f{1}{2}\pa{\ba{c}   2i\rho_{Z_n-\tf{\De c}{4}}  \\ \sqrt{c} \, \chi_n^{*}\ea} \;.
\enq
It is a reverse projector  at the points $\nu_n+ic, \ov{\nu}_n$
\beq
\pac{L_{0n}\pa{\nu_n+ic}}_{ab}= \de_b^{\pa{+}}\!\pa{n} \, \ga_a^{\pa{+}}\!\pa{n} \; , \quad \e{with} \qquad   
 \de^{\pa{+}}\!\pa{n} = \f{1}{2}\pa{\ba{c} \sqrt{c} \, \chi_n \\ -2i\rho_{Z_n-\tf{\De c}{4}} \ea} \; , \quad
\ga^{\pa{+}}\!\pa{n} = \pa{\ba{c} \sqrt{c} \, \chi_n^{*} \\ 2i\rho_{Z_n-\tf{\De c}{4}}  \ea} \;,
\enq
\beq
\pac{L_{0n}\pa{ \ov{\nu}_n } }_{ab}= \de_b^{\pa{-}}\!\pa{n} \, \ga_a^{\pa{-}}\!\pa{n} \; , \quad \e{with} \qquad  
 \de^{\pa{-}}\!\pa{n}=\pa{\ba{c} 2i\rho_{Z_n} \\ \sqrt{c} \, \chi_n^{*} \ea} \; , \quad
\ga^{\pa{-}}\!\pa{n}=\f{1}{2}\pa{\ba{c}   -2i\rho_{Z_n}  \\ \sqrt{c} \,  \chi_n \ea} \;.
\enq

\subsection{The lattice and the continuous models}

The Hamiltonian for the lattice model on an even number of sites $M$ is is built out of the monodromy matrix: 
\beq
T_{0;1\dots M}\!\pa{\la}\equiv T_{0}\!\pa{\la} = L_{0M}\pa{\la} \dots L_{01}\pa{\la} = 
\pa{ \ba{cc} A\pa{\la} & B\pa{\la} \\ C\pa{\la} & D\pa{\la} \ea }
\quad \e{with} \quad \;  M \in 2 \mathbb{Z} \;.
\enq
We have represented it as a $2\times 2$ matrix on the auxiliary space $V_0$
whose entries are operators acting on the quantum space $\mc{H}=\otimes_{n=1}^{M} \mc{H}_n$.
In the following, we set
\beq
\nu  \equiv \nu_{2n-1} = \nu_{2n} +i c = -\f{2i}{\De} +i \f{c}{2} \;.
\label{definition nu}
\enq
The fact that Lax matrices become projectors (or reverse projectors) at $\la=\nu$ allows one to build the below
local Hamiltonian out of the transfer matrix $\tau\pa{\la} = \e{tr}_0\pac{T_0\,\pa{\la}}$:
\bem
\tau^{-1}\!\pa{\nu} \cdot \tau^{\prime}\!\pa{\nu} = \sul{k=1}{M/2} 
\paa{ \, ^{t_0} \big[\be^{\pa{+}}\!\pa{2k+1}\!\big] L_{02k}\!\pa{\nu}L_{02k-1}\!\pa{\nu} \ga^{\pa{+}}\!\pa{2k-2} }^{-1}  \\
\cdot
\, ^{t_0}\big[\be^{\pa{+}}\!\pa{2k+1}\!\big] 
\Dp{\la}\pac{L_{02k}\!\pa{\la}L_{02k-1}\!\pa{\la}}^{\prime}_{\mid\la=\nu } \ga^{\pa{+}}\!\pa{2k-2}
\nonumber
\end{multline}
Above, $^{t_0}$ refers to the operation of transpotion of the vector $\be^{\pa{+}}\!\pa{2k+1}$. 
According to Izergin and Korepin \cite{IzerginKorepinLatticeVersionsofQFTModelsABANLSEandSineGordon},
the above local Hamiltonians goes, in the continuum limit:
\beq
\De \tend 0 \quad \e{with} \quad L=\De M \quad \e{fixed}
\enq
 to the Hamiltonian of the NLSM 
\beq
\bs{H}_{NLS} =\Int{0}{L} \paa{ \Dp{y} \Phi^{\dagger}\!\pa{y} \Dp{y} \Phi\pa{y} + c \, \Phi^{\dagger}\!\pa{y}\Phi^{\dagger}\!\pa{y} \Phi\pa{y} \Phi\pa{y}  } \dd y\; .
\label{definition Hamiltonien Bose}
\enq
In \eqref{definition Hamiltonien Bose} $\Phi$ and $\Phi^{\dagger}$ are canonical Bose fields subject to $L$ periodic boundary conditions. 
In such a continuous limit, the $k^{\e{th}}$ site of the lattice model can be though of as contributing to the "continuous coordinate" 
$x_k=k\De$. Then, the discreet fields $\chi_n$ are expected to be related to the canonical Bose fields $\Phi\pa{x}$ as
\beq
\chi_n=\Int{ n\De }{\pa{n+1}\De} \!\!  \Phi\pa{x}\;  \dd x \; . 
\enq
However, such an identification can only be given a formal sense in as much as, strictly speaking, the \textit{rhs} 
does not have a precise mathematical meaning. On the other hand, the \textit{lhs} has a sens in its own:
the local operators $\chi_n$ and $\chi^{*}_n$ can be constructed explicitly, for instance as the harmonic oscillator 
creation/annihilation operators. 

\subsection{The spectrum and eigenvectors}
\label{SousSectionSpectrumEigenvectors}

The transfer matrix $\la  \mapsto \tau\pa{\la}$ is diagonalized by means of standard considerations of the  algebraic
Bethe Ansatz.  One introduces the so-called pseudo-vacuum state $\ket{0}=\ket{0}_1 \otimes \dots \ket{0}_M $
where $\ket{0}_n$ is uniquely defined by the condition $\chi_n \ket{0}_n=0$ for all $n$. The commutation relations issuing 
from the Yang-Baxter equation \eqref{ecriture YBE} 
\beqa
A\pa{\la} B\pa{\mu} &=& \f{\la-\mu+ic}{\la-\mu} B\pa{\mu} A\pa{\la}  - \f{ic}{\la-\mu}  B\pa{\la} A\pa{\mu} \\
D\pa{\la} B\pa{\mu} &=& \f{\la-\mu-ic}{\la-\mu} B\pa{\mu} D\pa{\la}  + \f{ic}{\la-\mu}  B\pa{\la} D\pa{\mu} \; , 
\eeqa
lead to the conclusion that the state
\beq
\big| \, \psi \, \big( \! \paa{ \la_a }_1^N \! \big) \big> =B\pa{ \la_1}\dots B\pa{\la_N} \ket{0}
\enq
is an eigenstate of the transfer matrix $\tau\!\pa{\la}$  associated with the eigenvalue
\beq
\La\pa{\la\mid\paa{\la}_1^N} = a\pa{\la} \pl{p=1}{N} \f{\la-\la_p+ic}{\la-\la_p} \; + \;  d\pa{\la} \pl{p=1}{N} \f{\la-\la_p-ic}{\la-\la_p}
\enq
where
\beq
a\pa{\la} = \paa{ -i \f{ \la \De}{2} +1 + \f{c\De}{4}  }^{\f{M}{2}} \hspace{-2mm} \cdot \, \paa{ -i \f{ \la \De}{2} +1 - \f{c\De}{4}  }^{\f{M}{2}}
\quad \e{and} \quad
d\pa{\la} = \paa{ i \f{ \la \De}{2} +1 + \f{c\De}{4}  }^{\f{M}{2}} \hspace{-2mm}  \cdot \, \paa{ i \f{ \la \De}{2} +1 - \f{c\De}{4}  }^{\f{M}{2}} \; 
\enq
provided that the parameters $\paa{\la_a}_1^N$ solve the Bethe Ansatz equations (BAE)
\beq
\f{d\pa{\la_r}}{a\pa{\la_r}} = \pl{ \substack{p=1 \\ p \not= r} }{N} \f{ \la_r-\la_p + ic  }{ \la_r-\la_p - ic } \;\;\; , \qquad 
r=1,\dots,N \;.
\label{ecriture BAE modele discret}
\enq
The solutions to \eqref{ecriture BAE modele discret} are real valued, satisfy to the so-called repulsion principle:
\beq
 \e{if} \; a \not= b \quad \e{then} \quad \la_a \not= \la_b \;, 
\label{repulsion principle}
\enq
and are in a one-to-one correspondence with a certain subset (depending on $\De$ and $L$ for $\De M=L$ fixed) 
of the sets of all ordered integers $\ell_1 < \dots < \ell_{N}$, $\ell_a \in \mathbb{Z}$. 
More precisely, given any choice of integers $\ell_1<\dots <\ell_N$, 
there exists a $\wt{\De}$ such that, for $\De<\wt{\De}$ (with $\De M=L$ fixed) there exists a unique solution to the below set of logarithmic Bethe equations
\beq
-i\ln\paf{ d \pa{ \mu_{\ell_r} }  }{ a\pa{ \mu_{\ell_r} }  } \; +  \;
\sul{p=1}{N} \th \pa{\mu_{\ell_r} - \mu_{\ell_p} } = 2\pi \pa{ \ell_r - \f{N+1}{2} } \;\;\; , \quad 
r=1,\dots,N \; \; \; \e{with} \;\;\th\pa{\la} = i  \ln  \paf{ ic + \la  }{ ic - \la  } \;.
\label{ecriture log BAE}
\enq
Finally, using elementary properties of \eqref{ecriture log BAE}, it can be shown that, 
given a fixed product $\De M =L$ and any choice of inegers $\ell_1<\dots <\ell_N$, there exists a $\De_0>0$ such that the parameters 
$\mu_{\ell_a}=\mu_{\ell_a}\pa{\De}$ are continuous in $\De\in \intff{0}{\De_0}$. 

In fact, the $\De \tend 0$, $M\De =L$ limit of such a solution $\mu_{\ell_a}^{\e{c}}=\lim_{\De \tend 0} \mu_{\ell_a}\pa{\De} $
gives rise to the set of parameters solving the logarithmic Bethe equations arizing in the $N$ quasi-particle sector of the continuous model described by 
the Hamiltonian \eqref{definition Hamiltonien Bose}:
\beq
L \mu_{\ell_r}^{\e{c}} \; +  \;
\sul{p=1}{N} \th \pa{\mu^{\e{c}}_{\ell_r} - \mu^{\e{c}}_{\ell_p} } = 2\pi \pa{ \ell_r - \f{N+1}{2} } \;\;\;, \qquad 
r=1,\dots,N \;.
\label{ecriture Log BAE continuous model}
\enq
Throughout this paper, we will always use the superscript $\e{c}$ so as to indicate that $\{ \mu_{\ell_a}^{\e{c}}\}_1^N$
stands for the solution of the Bethe Ansatz equations for the continous model. Likewise, the absence of such a superscript will
indicate that one deals with the solution of the model at finite $\De$. We will omit the explicit wirting of this $\De$
dependence. 

It has been shown in \cite{DorlasOrthogonalityAndCompletenessNLSE} that the vectors $\ket{ \psi \, \big( \! \paa{ \mu_{\ell_a} }_1^N \! \big) }$
converge, in some suitable sense, to the eigenfunctions 
\beq
\big| \, \Psi\big(  \{ \mu_{\ell_a}^{\e{c}} \}_{a=1}^{N} \, \big)  \big> = \Int{0}{L} 
\vp\big( x_1,\dots, x_N \mid \{ \mu_{\ell_a}^{\e{c}} \}_1^N \big)  \; 
\Phi^{\dagger}\pa{x_1} \dots  \Phi^{\dagger}\pa{x_N} \ket{0}  \,  \dd^N\!x \;\;  
\enq
of the continuous Hamiltonian \eqref{definition Hamiltonien Bose} in the $N$ quasi-particle sector. 
The function $\vp\pa{ x_1,\dots, x_N \mid \{ \la_a \}_1^N }$ can be constructed by 
means of the coordinate Bethe Ansatz \cite{LiebLinigerCBAForDeltaBoseGas} and reads
\beq
\vp\big( x_1, \dots, x_N  \mid \{ \la_a \}_1^N \big) = \pa{-i \sqrt c}^{N}
\sul{ \sg \in \mf{S}_N }{} \pl{ a<b }{ N } \paa{ \f{ \la_{\sg\pa{a}} - \la_{\sg\pa{b}} -i c \e{sgn}\pa{x_a-x_b}  }
{  \la_{\sg\pa{a}} - \la_{\sg\pa{b}} } } \cdot 
\pl{a=1}{N} \ex{i\la_{\sg\pa{a} } x_a } \ex{-i \la_{\sg\pa{a}} \f{L}{2} } \;.
\label{ecriture fonction propre model continu}
\enq
In \eqref{ecriture fonction propre model continu} we made use of the following definition for the sign function:
\beq
\e{sgn}\pa{x} =  1 \quad \e{for} \; x >0 \qquad , \quad \e{sgn}\pa{x} =  0 \quad \e{for} \; x =0 \qquad , \quad 
\e{sgn}\pa{x} =  -1 \quad \e{for} \; x <0 \;.
\label{definition fonction signe}
\enq

\subsection{Structure of the space of states}

The very setting of the algebraic Bethe Ansatz allows one to characterize the structure of the space of states by providing
determinants representations for the norms \cite{KorepinNormBetheStates6-Vertex} and the scalar products between Bethe vectors 
\cite{SlavnovScalarProductsXXZ}. 
\begin{prop} \cite{KorepinNormBetheStates6-Vertex}
\label{Proposition normes etat Bethe}
Let $\paa{\mu_{\ell_a}}_1^{N+1}$ be any solution to the Bethe Ansatz equations \eqref{ecriture log BAE},    then the norm of the 
associated Bethe state admits the below determinant representation
\beq
\norm{\psi\pa{\paa{\mu_{\ell_a}}_1^{N+1} }}^2 = 
\pl{a=1}{N+1} \paa{ 2i\pi L \wh{\xi}^{\prime}_{\paa{\ell_a}}\!\pa{\mu_{\ell_a}} a(\mu_{\ell_a}) d(\mu_{\ell_a})}
\f{ \pl{a,b=1}{N+1} \pa{\mu_{\ell_a}-\mu_{\ell_b}-ic }  }{ \pl{ \substack{ a,b=1 \\ a\not= b} }{N+1} \pa{\mu_{\ell_a}-\mu_{\ell_b}  }  } 
\ddet{N+1}{ \Xi^{\pa{\mu}}} \;.
\enq
The entries of the matrix $\Xi^{\pa{\mu}}$ read 
\beq
\Xi^{\pa{\mu}}_{jk} = \de_{jk} -\f{  K\pa{\mu_{\ell_a}-\mu_{\ell_b}} }{ 2\pi L \wh{\xi}_{\paa{\ell_a}}^{\prime}\!\pa{\mu_{\ell_b}} }
\quad \e{with} \quad 
\wh{\xi}_{\{\ell_a\}}\pa{\om} = -\f{i}{2\pi L} \ln\paf{ d \pa{ \om }  }{ a\pa{ \om }  } \; +  \;
\f{1}{2\pi L}\sul{p=1}{N+1} \th \big(\om - \mu_{\ell_p} \big)   \;+\; \f{N+2}{2 L} \; ,
\enq
and we have agreed upon $K\pa{\la}=\th^{\prime}\!\pa{\la}$. 
\end{prop}

\begin{theorem} \cite{SlavnovScalarProductsXXZ}
\label{Theorem Nikita Scalar Products}
Let $\{ \mu_{\ell_a} \}_1^{N+1}$ be a solution to the logarithmic Bethe equations \eqref{ecriture log BAE}
and $\paa{ \la_a }_{1}^{N+1}$ a generic set of parameters. Then, the below scalar product reads
\beq
\braket{  \psi\pa{ \{ \mu_{\ell_a} \}_{1}^{N+1} }  }{ \psi\pa{ \{\la_a \}_{1}^{N+1} } } = 
\f{ \pl{a=1}{N+1} d\pa{\mu_{\ell_a} } }{ \pl{a>b}{N+1} \pa{\mu_{\ell_a}-\mu_{\ell_b}}\pa{\la_{b}-\la_{a}} }
\ddet{N+1}{ \Om\pa{  \{ \mu_{\ell_a} \} , \{ \la_a \}  }  } \;,
\label{ecriture produits scalaires}
\enq
where
\beq
\pac{ \Om\pa{  \{ \mu_a \} , \{ \la_a \}  } }_{jk}=
a\pa{\la_k} t\pa{\mu_j,\la_k} \pl{a=1}{N+1} \pa{\mu_a-\la_k-ic} \; -\; d\pa{\la_k} t\pa{\la_k,\mu_j} \pl{a=1}{N+1} \pa{\mu_a-\la_k+ic}
\enq
and
\beq
t\pa{\la,\mu}  = \f{ -ic  }{ \pa{\la-\mu}\pa{\la-\mu-ic}} \;.
\enq
\end{theorem}

It was found by Oota \cite{OotaInverseProblemForFieldTheoriesIntegrability} that the reduction of the 
Lax matrix to projectors at zeroes of the quantum determinant that allows one to build local Hamiltonians
from the transfer matrix can also be used to reconstruct certain local operators of the theory. 
In particular, one has the identity 
\beq
\tau^{-1}\!\! \pa{\nu} \cdot  B\pa{\nu}  = \bigg\{ \sul{r=1}{2} \ga_r^{\pa{+}}\!\pa{M}\be_r^{\pa{+}}\!\pa{1}  \bigg\}^{-1} \hspace{-2mm} \cdot  
\, \ga_{1}^{\pa{+}}\!\pa{M} \be_2^{\pa{+}}\!\pa{1} \; .
\label{ecriture reconstruction Oota}
\enq
Using the explicit formulae for $\ga^{\pa{+}}\pa{k}$ $\be^{\pa{+}}\pa{k}$ one gets 
\beq
\sul{r=1}{2}\ga_r^{\pa{+}}\!\pa{M} \be_r^{\pa{+}}\!\pa{1} = \f{c}{2} \, \chi_M^{*} \, \chi_1 \; + \;  2 \rho_{Z_M-\f{\De c}{4}} \, \rho_{Z_1}
\quad \e{and} \quad \ga_1^{\pa{+}}\!\pa{M}\be_2^{\pa{+}}\!\pa{1}  =  
-i\sqrt{c} \, \chi_M^{*} \, \rho_{Z_1} \;.
\enq
Thus, at least formally, one expects the below reconstruction formula for operators in the continuous model to hold.
\beq
\tau^{-1}\!\! \pa{\nu} \cdot   B\pa{\nu} =  -\f{i\sqrt{c}}{2} \De \, \Phi^{\dagger}\! \pa{0} +\e{O}\,\big( \De^2 \big) \;.
\label{ecriture identification formelle ac champ continu}
\enq

\section{Form factors of the conjugated field operator}
\label{Section FF resultat principal}

The formal identification \eqref{ecriture identification formelle ac champ continu} of products of entries
of the monodromy matrix with operators in the continuous model can be made rigorous. 
This is one of the main results of this paper. It allows one to provide the missing steps in the passage 
from determinant representations for certain local operators in the lattice model obtained through the solution of the inverse problem 
\cite{OotaInverseProblemForFieldTheoriesIntegrability} to those for the 
form factors of the local operators in the continuous case. 
The proof of this theorem is postponed to appendix \ref{Appendix Proof continuous limit}. 
There, we also prove a similar result for the determinant representations of scalar products
for the continuous model.  
\begin{theorem}
Let $\{ \la_{\ell_a} \}_1^{N}$ be a solution of the logarithmic Bethe equations  \eqref{ecriture log BAE}
in the $N$ particle sector and  $\{\mu_a\}_1^N$ a set of generic, pairwise distinct complex numbers.  Then 
the below scalar product in the lattice model converges, in the $\De \tend 0$ limit, to the scalar product in the continuous model
\beq
\braket{  \psi\pa{ \{ \mu_{a} \}_{1}^{N} }  }{ \psi\pa{ \{\la_{\ell_a} \}_{1}^{N} } } \limit{ \De }{ 0 }
 \Int{0}{L} \f{ \dd ^N x }{ N! } \;  \ov{ \vp\big( x_1, \dots, x_N \mid \{ \mu_{a} \}_1^{N} \big) } \; 
\vp\big( x_1,\dots, x_N \mid  \{ \la_{\ell_a}^{\e{c}} \}_1^N \big) \;.
\label{equation convergence PS discret vers integrale}
\enq
As a consequence, one has the below determinant representation for the scalar products in the continuous model:
\beq
\f{ \pl{a=1}{N} d\big(\la_{\ell_a}^{\e{c}} \big) }{ \pl{a>b}{N} \big(\la^{\e{c}}_{\ell_a}-\la^{\e{c}}_{\ell_b}\big)\pa{\mu_{b}-\mu_{a}} }
\ddet{N+1}{ \Om\big(  \{ \la^{\e{c}}_{\ell_a} \}_1^N , \{ \mu_a \}_1^N  \big)  } \;. 
\enq
\end{theorem}
\begin{theorem}
\label{Theorem cvgce lattice discreization}
Let $\{ \mu_{\ell_a} \}_1^{N+1}$ and $\{\la_{r_a}\}_1^N$ be any two solution of the logarithmic Bethe equations  \eqref{ecriture log BAE}
in the $N+1$ and $N$ particle sectors respectively. Then, the expectation value 
\beq
F_{\Phi^{\dagger}}^{\pa{\De}}\pa{  \{ \mu_{\ell_a} \}_1^{N+1} ;  \{ \la_{r_a} \}_1^N } =  \f{ 2 i }{\De \sqrt{c}}
\cdot  \bra{ \psi\pa{ \{ \mu_{\ell_a} \}_1^{N+1} } }  \, \tau^{-1}\! \pa{\nu} B\pa{\nu} \, \ket{ \psi\pa{ \{ \la_{r_a} \}_1^N} } \;, 
\label{definition valeur moyenne discrete pour FF champ conj}
\enq
converges to the below form factor of the field operator in the continuous model
\beq
F_{\Phi^{\dagger}}\pa{  \{ \mu_{\ell_a}^{\e{c}} \}_1^{N+1} ;  \{ \la_{r_a}^{\e{c}} \}_1^N } =
 \Int{0}{L} \f{ \dd ^N x }{ N! } \;   \ov{ \vp\big( 0,x_1, \dots, x_N \mid \{ \mu_{\ell_a}^{\e{c}} \}_1^{N+1} \big) } \cdot 
\vp\big( x_1,\dots, x_N \mid  \{ \la_{r_a}^{\e{c}} \}_1^N \big) \;.
\enq
The latter admits the below determinant representation 
\beq
F_{\Phi^{\dagger}}\pa{ \{ \mu_{\ell_a}^{\e{c}} \}_1^{N+1}; \paa{\la^{\e{c}}_{r_a} }_1^N} = %\lim_{\De \tend 0} \f{- 2 }{i\De \sqrt{c}}
%
%\bra{\paa{\mu}_1^{N+1}} \tau^{-1}\pa{\nu} B\pa{\nu} \ket{\paa{\la}_1^N} \\
%
%
i \sqrt{c} \pl{a=1}{N+1} \ex{\f{i L }{2} \mu_{\ell_a}^{\e{c}} } \cdot
\pl{k=1}{N} \paa{ \ex{-\f{iL}{2}\la_{r_k}^{\e{c}} }\pac{1-\ex{-2i\pi \wh{F}^{\e{c}}( \la_{r_k}^{\e{c}} )  }} 
\pl{b=1}{N+1} \f{\mu_{\ell_b}^{\e{c}}-\la_{r_k}^{\e{c}}-ic}{\mu_{\ell_b}^{\e{c}}-\la_{r_k}^{\e{c}} } }
\ddet{N}{\de_{jk} + U_{jk} } \;.
\enq
%
%
%

%There $ \msc{C}_q $ in a small counterclockwise loop around $\intff{-q}{q}$ such that it avoids all zeroes of  
%$\ex{-2i\pi \wh{F}^{\e{c}}_{\paa{\ell_a}}\pa{\om}} -1$ located in this neighborhood. The kernel of the integral operator reads
%
%
%
\beq
U_{jk} = - i  \pl{a=1}{N+1}\f{ \la_{r_j}^{\e{c}}-\mu_{\ell_a}^{\e{c}} }{ \la_{r_j}^{\e{c}} - \mu_{\ell_a}^{\e{c}} +ic} \cdot 
 \f{ \pl{a=1}{N} \pa{ \la_{r_j}^{\e{c}} -\la_{r_a}^{\e{c}}+ic } }{ \pl{ \substack{a=1 \\ \not=j } }{N} \pa{\la_{r_j}^{\e{c}}-\la_{r_a}^{\e{c}} }}  
 \cdot \f{ K \pa{  \la_{r_j}^{\e{c}}- \la_{r_k}^{\e{c}} }  }{ \ex{-2i\pi \wh{F}^{\e{c}}( \la_{r_j}^{\e{c}}) } -1 } \;.
\enq
Above, we made use of the so-called the discreet shift function $\wh{F}^{\e{c}}$ for the continuous model:
\beq
\ex{-2i\pi \wh{F}^{\e{c}}\pa{\om}} = \pl{a=1}{N+1} \f{ \mu^{\e{c}}_{\ell_a}-\om + ic }{ \mu^{\e{c}}_{\ell_a}-\om - ic } \cdot
\pl{a=1}{N} \f{ \la^{\e{c}}_{r_a}-\om - ic }{ \la^{\e{c}}_{r_a}-\om + ic } \;.
\enq
\end{theorem}

\subsection{Determinant representation in the lattice model}

Determinant representations for the form factors of the conjugated field operator in the NLSM
have been obtained in \cite{KojimaKorepinSlavnovNLSESystemPartialDiffEqnsDualFieldTempeAndTime} through the use of the two-site model, 
and in \cite{OotaInverseProblemForFieldTheoriesIntegrability} with the help of the inverse problem previously discussed. 
These results all relied on the hypothesis of the convergence of the lattice discretiztion to the continuous model 
has been proven in theorem \ref{Theorem cvgce lattice discreization} above. 
Actually, we have provided a slightly different (in respect to the aforecited 
\cite{KojimaKorepinSlavnovNLSESystemPartialDiffEqnsDualFieldTempeAndTime,OotaInverseProblemForFieldTheoriesIntegrability}) 
determinant representation for $F_{\Phi^{\dagger}}\pa{  \{ \mu_{\ell_a}^{\e{c}} \}_1^{N+1} ;  \{ \la_{r_a}^{\e{c}} \}_1^N } $. 
The equivalence of our represention with the previous ones can, in principle, be checked with the help of 
determinant identities analogous to those established in 
\cite{KozKitMailSlaTerXXZsgZsgZAsymptotics,KozKitMailSlaTerThermoLimPartHoleFormFactorsForXXZ}. 
We now derive a determinant representation for $F_{\Phi^{\dagger}}^{\pa{\De}}$ defined in \eqref{definition valeur moyenne discrete pour FF champ conj}.
This provide a slightly different representation in respect to the one obtained by Oota \cite{OotaInverseProblemForFieldTheoriesIntegrability}.
\begin{prop}
\label{Proposition facteur de forme Psi}
The discreet approximation 
$F_{\Phi^{\dagger}}^{\pa{\De}}\pa{  \{ \mu_{\ell_a} \}_1^{N+1} ;  \{ \la_{r_a} \}_1^N }$ defined in 
\eqref{definition valeur moyenne discrete pour FF champ conj} admits the determinant representation
\bem
F_{\Phi^{\dagger}}^{\pa{\De}}\pa{\paa{ \mu_{\ell_a} }_1^{N+1}; \paa{ \la_a }_1^N} =  \f{- 2 \sqrt{c} }{ \De  } 
 \pl{a=1}{N+1} \f{\nu-\mu_{\ell_a} }{\nu-\mu_{\ell_a}-ic} \f{ \pl{a=1}{N}\pa{\la_{r_a}-\nu + ic} }{ \pl{a=1}{N+1} \pa{\mu_{\ell_a}-\nu} }
\pl{a=1}{N+1}\pl{b=1}{N} \f{1}{\mu_{\ell_a}-\la_{r_b}  } \\
\pl{a=1}{N+1}d\pa{ \mu_{\ell_a} } \; \pl{k=1}{N} \paa{ a\pa{\la_{r_k}}\pac{ 1-\ex{-2i\pi \wh{F}(\la_{r_k} ) } } 
\pl{b=1}{N+1}\pa{\mu_{\ell_a}-\la_{r_k}-ic} }
\cdot \ddet{ N }{ \de_{jk}+ U^{\pa{\De}}_{jk}  }\;.
\end{multline}
%
%
%

%There $ \msc{C}_q $ in a small counterclockwise loop around $\intff{-q}{q}$ such that it avoids all zeroes of  
%$\ex{-2i\pi \wh{F}^{\e{c}}_{\paa{\ell_a}}\pa{\om}} -1$ located in this neighborhood. The kernel of the integral operator reads
%
%
%
%
%
%
\beq
U_{jk}^{\pa{\De}} = - i  \pl{a=1}{N+1}\f{ \la_{r_j}-\mu_{\ell_a} }{ \la_{r_j} - \mu_{\ell_a} +ic} \cdot 
 \f{ \pl{a=1}{N} \big( \la_{r_j} -\la_{r_a}+ic \big) }{ \pl{ \substack{a=1 \\ \not=j } }{N} \big( \la_{r_j}-\la_{r_a} \big) }  
 \cdot \f{ K \pa{  \la_{r_j}, \la_{r_k}  \mid \nu}  }{ \ex{-2i\pi \wh{F} ( \la_{r_j}) } -1 } \;.
\enq
and, recalling that $K\pa{\la}=\th^{\prime}\pa{\la}$ with $\th\pa{\la}$ given in \eqref{ecriture log BAE} and $\nu$ in \eqref{definition nu}, 
\beq
K\pa{\om,\om^{\prime}\mid \nu}  =  \f{\nu-\om-ic}{\nu-\om}
\paa{  K\pa{\om-\om^{\prime}}  \; -i \; \pa{1 - \f{\nu-\om^{\prime}+ic}{\nu-\om^{\prime}-ic } } 
\pa{  \f{1}{\om-\om^{\prime} +ic} -\f{1}{\om-\nu+ic} } } \;.
\enq
Also, above, we made use of the so-called the discreet shift function $\wh{F}^{\e{c}}$ for the continuous model:
\beq
\ex{-2i\pi \wh{F}\pa{\om}} = \pl{a=1}{N+1} \f{ \mu_{\ell_a}-\om + ic }{ \mu_{\ell_a}-\om - ic } \cdot
\pl{a=1}{N} \f{ \la_{r_a}-\om - ic }{ \la_{r_a}-\om + ic } \;.
\label{definition fonction shift}
\enq

\end{prop}

\Proof

Using that $\ket{ \psi\pa{\paa{\mu_{\ell_a}}_1^{N+1} } }$ is an eigenstate of $\tau^{-1}\!\pa{\la}$ for any $\la$, 
it readily follows that 
\beq
F_{\Phi^{\dagger}}^{\pa{\De}}\pa{\paa{\mu_{\ell_a}}_1^{N+1}; \paa{\la_{r_a}}_1^N} =  \f{ 2 i }{\De \sqrt{c}} \pac{ d\pa{\nu}}^{-1}
\pl{a=1}{N+1} \f{\nu-\mu_{\ell_a}}{\nu-\mu_{\ell_a}-ic} 
\cdot \braket{ \psi\pa{\paa{\mu_{\ell_a}}_{1}^{N+1}} }{ \psi\pa{\paa{\la_{r_a}}_{1}^{N+1}} } \;.
\enq
In the scalar product formula, we agree upon $\la_{r_{N+1}}\equiv \nu$. 

Using techiques proposed in \cite{KozKitMailSlaTerThermoLimPartHoleFormFactorsForXXZ,KozKitMailSlaTerXXZsgZsgZAsymptotics}
it is possible to factor out a Cauchy determinant from the determinant of $\Om$. This leads to the representation
\beq
\ddet{N+1}{ \Om\pa{\paa{\mu_{\ell_a}} , \paa{\la_{r_a} } } } = \ddet{N+1}{\f{1}{\mu_{\ell_a}-\la_{r_b} }} \cdot \ddet{N+1}{S} \;.
\enq
The matrix $S$ takes the form
\beq
S_{jk}= \de_{jk} \mc{Y}\pa{\la_{r_k}\mid\paa{\mu_{\ell_a}}_1^{N+1} } \; + \;  
\f{ \pl{a=1}{N+1} \pa{\la_{r_j}-\mu_{\ell_a} } }{  \pl{\substack{a=1\\ a\not= j} }{N+1} \pa{\la_{r_j}-\la_{r_a} }  }
\cdot \f{\Dp{}}{\Dp{}y_j} \mc{Y}\pa{\la_k\mid\paa{y_a}} \mid_{\paa{y_a}=\{ \la_{r_a} \} }
\enq
for $k\in\intn{1}{N+1}$, $j\in\intn{1}{N}$ and  $S_{N+1 \, k}= \mc{Y}\pa{\la_{r_k}\mid\paa{\la_{r_a}}}$.

Above, we have set
\beq
\mc{Y}\pa{\la\mid\paa{\tau}_1^{N+1}}= a\pa{\la} \pl{k=1}{N+1} \pa{\tau_k-\la - ic} +  d\pa{\la} \pl{k=1}{N+1} \pa{\tau_k-\la + ic} \;.
\enq
One can reduce the dimensionality of $\ddet{N+1}{S}$ by $1$ thanks to the below linear combination of columns
\beq
C_k \leftarrow C_k-  \f{ \mc{Y}\pa{\la_{r_k} \mid \paa{ \la_{r_a} }_1^{N+1} } }{ \mc{Y}\pa{ \nu \mid \paa{ \la_{r_a} }_1^{N+1}} } \cdot C_{N+1} \;.
\enq
Then, using explicitly that $\la_{r_{N+1}}=\nu$, one gets 
\beq
\ddet{N+1}{S}= \mc{Y}\pa{\nu\mid\paa{\la_{r_a} }_1^{N+1}} \cdot
 \ddet{N}{ S_{jk}- S_{j\, N+1} \f{ \mc{Y}\pa{\la_{r_k} \mid \paa{ \la_{r_a} }_1^{N+1} } }{ \mc{Y}\pa{\nu \mid \paa{ \la_{r_a} }_1^{N+1}} }  } \;.
\enq

The functions $\mc{Y}\pa{\la_{r_k}\mid\paa{\mu_{\ell_a}}_1^{N+1}}$ can be recast in terms of the shift function $\wh{F}\!\pa{\la_{r_k}}$
given in \eqref{definition fonction shift}
\beq
\mc{Y}\pa{\la_{r_k}\mid\paa{\mu_{\ell_a}}_1^{N+1}} = a\pa{\la_k} \pl{a=1}{N+1} \pa{\mu_{\ell_a}-\la_{r_k}-ic} \cdot \paa{ 1-\ex{-2i\pi \wh{F}\pa{\la_{r_k}}} } \;.
\label{equation factorisation alternative Y kappa}
\enq
To obtain \eqref{equation factorisation alternative Y kappa} we have used that $\paa{\la_{r_k}}_{k=1}^{N}$ is
the solution of the $N$-particle BAE. Then, computing explicitly the difference in the determinant and factoring out the $\mc{Y}$ functions,
we get that
\beq
\ddet{N+1}{ \Om\pa{\paa{\mu_{\ell_a}} , \paa{\la_{r_a} } } } 
= \mc{Y}\pa{\nu\mid\paa{\la_{r_a}}_1^{N+1}} \cdot  \pl{k=1}{N} \mc{Y}\pa{\la_{r_k}\mid\paa{\mu_{\ell_a}}_1^{N+1}} \cdot
\ddet{N+1}{\f{1}{\mu_{\ell_a}-\la_{r_a} } }  \ddet{N}{\de_{jk}+ U^{\pa{\De}}_{jk}  } \;.
\enq
It only remains to put all the formulae together. \qed

\section{Large volume behavior of the Form Factors of conjugated fields}
\label{Section Thermo limit FF}

In this subsection, we provide formulae for the large volume limit of the form factors $F_{\Phi^{\dagger}}$
for a specific class of excited states.  Namely, we assume that the state described by $\{\mu_{\ell_a}^{\e{c}}\}_1^{N+1}$
correponds to an $n$-particle/hole excitation above the $N+1$-quasi particle ground state whereas the state $\{\la_{r_a}^{\e{c}} \}_1^N$
stands for the ground state (\textit{ie} $r_a=a$ for $a=1,\dots,N$) in the $N$-quasi particle sector.
The methods for carrying out such computations have been developped in 
\cite{KozKitMailSlaTerThermoLimPartHoleFormFactorsForXXZ,KozKitMailSlaTerEffectiveFormFactorsForXXZ,SlavnovFormFactorsNLSE}.

\subsection{Rudiemnts of the thermodynamic limit in the NLSM}

Given the set of Bethe roots $\{\la_a^{\e{c}}\}$ for the ground state in the $N$ quasi-particle sector, one builds  their 
counting function as
\beq
\wh{\xi}\pa{\om} \equiv \wh{\xi} \pa{\om \mid \paa{\la_{a}^{\e{c}}}_{1}^{N} } = \f{\om}{2\pi} \; + \;
 \f{1}{2\pi L} \sul{a=1}{ N } \th\pa{  \om-\la_a^{\e{c}} }
\; +  \; \f{N+1}{2L} \;, \quad  ie \;\; \wh{\xi}\pa{\la_a^{\e{c}}}=\f{a}{L} \;. 
\label{definition counting function la}
\enq
The latter has the below behavior in the thermodynamic limit of the model ( \textit{ie} $N, L\tend +\infty$ with $N/L\tend D$)
\beq
\wh{\xi}\pa{\om } = \xi\pa{\om} + \e{O} ( L^{-1} ) \quad \e{where} \quad
\xi\pa{\om}= \f{p\pa{\om}}{2\pi} + \f{D}{2} \qquad \e{with} \quad  
p\pa{\la} - \Int{-q}{q} \th\pa{\la-\mu} p^{\prime}\pa{\mu} \f{\dd \mu}{2\pi} = \la \:.
\label{ecriture limite thermo fction comptage}
\enq
The parameter $q$ corresponds to the right end of the Fermi interval $\intff{-q}{q}$ on which the ground state's Bethe roots condensate
in the thermodynamic limit. It is defined as the unique solution to $ p\pa{q}= \pi D$.

Recall that  any solution $\{\mu^{\e{c}}_{\ell_a}\}_1^{N+1}$ of the Bethe equations in the $N+1$ quasi-particle sector
is uniquely detemined by the choice of $N+1$ integers $\ell_1< \dots < \ell_{N+1} $. 
It is convenient to parameterize the integers $\ell_j$ in terms of particle-hole excitations above the
$N+1$ quasi-particle ground state:
\beq
\ell_j=j \quad \e{for} \; \; j \in \intn{1}{N +1 } \setminus{h_1,\dots,h_n}  \quad \e{and}  \quad
\ell_{h_a}=p_a \quad \e{for} \;\;  a=1,\dots,n \; .
\label{definition correpondance entiers ella et particules-trous}
\enq
The integers $p_a$ and $h_a$ are such that $p_a \not \in \intn{1}{N +1 }\equiv \{ 1,\dots, N +1  \}$ and $h_a\in \intn{1}{N +1 }$.

One can actually associate a counting function to any solution $\{\mu^{\e{c}}_{\ell_a}\}_1^{N+1}$ by
\beq
\wh{\xi}_{\paa{\ell_a}} \pa{\om} \equiv \wh{\xi}_{\paa{\ell_a}} \big( \om \mid \{ \mu_{\ell_a}^{\e{c}} \}_{1}^{N+1} \big) = 
\f{\om}{2\pi} +\f{1}{2\pi L} \sul{a=1}{N+1} \th\big(\om-\mu^{\e{c}}_{\ell_a} \big) + \f{N+2}{2L} \;.
\label{definition counting function mu}
\enq
By construction, it is such that $\wh{\xi}_{\paa{\ell_a}} \big(\mu^{\e{c}}_{\ell_a} \big)=\tf{\ell_a}{L}$, for $a=1,\dots, N + 1$.
Actually, $\wh{\xi}_{\paa{\ell_a}}\pa{\om }$ defines a set of background parameters $\paa{\wh{\mu}_a}$, $a\in \mathbb{Z}$,
as the unique solutions to $\wh{\xi}_{\paa{\ell_a}}\pa{\wh{\mu}_a}=\tf{a}{L}$. The latter allows one to define the rapidities
$\wh{\mu}_{p_a}$, resp. $\wh{\mu}_{h_a}$, of the particles, resp. holes, entering in the description of 
$\{ \mu_{\ell_a}^{\e{c}} \}_1^{N +1 }$.

It can be shown that the shift function 
\beq
\wh{F}^{\e{c}}\pa{\om}= L \pac{\wh{\xi}\pa{\om}  - \wh{\xi}_{\paa{\ell_a}}\pa{\om} }
\enq
has a well defined thermodynamic limit
\beq
F\pa{\la} \equiv  F\pabb{\la}{ \{ \mu_{p_a} \} } {\paa{\mu_{h_a}} }
 =  -\tf{Z\pa{\la}}{2} \; -\; \phi\pa{\la,q} \; - \; \sul{a=1}{n} \pac{\phi ( \la,\mu_{p_a}) - \phi\pa{\la,\mu_{h_a}}}
\label{ecriture limite thermo fction shift}
\enq
where the dressed phase $\phi\pa{\la,\mu}$ and the dressed charge $Z\pa{\la}$ solve the linear integral equations
\beq
\phi\pa{\la,\mu}- \Int{-q}{q} K\pa{\la-\tau} \phi\pa{\tau,\mu} \f{\dd \tau}{2\pi} =  \f{1}{2\pi}\th\pa{\la-\mu} \qquad \e{and} \qquad
Z\pa{\la}- \Int{-q}{q} K\pa{\la-\tau} Z\pa{\tau} \f{\dd \tau}{2\pi} =  1 \;.
\label{definition eqn int Z et phi}
\enq
This thermodynamic limit of the shift function depends on the particles' $\{\mu_{p_a}\}$ and holes' $\{\mu_{h_a}\}$ positions in the
thermodynamic limit. These are defined as the solutions to 
\beq
\xi ( \mu_{p_a} ) =\tf{p_a}{L} \qquad \e{and} \qquad  \xi\pa{\mu_{h_a}}=\tf{h_a}{L} \; . 
\label{definition particle trou thermo lim}
\enq
We remind that the above shift function measures the spacing between the ground state roots $\la_a$ and the background parameters $\wh{\mu}_a$ defined by
$\wh{\xi}_{\paa{\ell_a}}$\,: \;  $\wh{\mu}_a-\la_a = F\pa{\la_a}\cdot \pac{L \xi^{\prime}\!\pa{\la_a}}^{-1} 
\big( 1+ \e{O}\big(L^{-1}\big) \big)$.

\subsection{Thermodynamic limit of form factors}
\label{Subsection Thermo Lim FF}

By applying propositions \ref{Proposition normes etat Bethe} and \ref{Proposition facteur de forme Psi}
%and utilizing the identity $F_{\Phi^{\dagger}}\pa{ \{ \mu_{\ell_a}^{\e{c}} \}_1^{N+1}; \{ \la^{\e{c}}_a \}_1^N }
%
%= F_{\Phi}^*\pa{ \{\la^{\e{c}}_a \}_1^N ;\{ \mu_{\ell_a} \}_1^{N+1}} $ 
it readily follows that the normalized modulus squared of the form factor of the conjugated field  admits the factorization 
\beq
\f{  \abs{ \bra{\Psi\pa{\{\mu_{\ell_a}^{\e{c}} \}_1^{N+1}}} \Phi^{\dagger}\pa{0} \ket{ \Psi\pa{ \{ \la_a^{\e{c}} \}_1^N}  } }^2 }
{ \norm{\Psi\pa{ \{ \mu^{\e{c}}_{\ell_a} \}_{1}^{N+1} }}^2  \norm{\Psi\pa{\{ \la^{\e{c}}_a \}_{1}^{N} }}^2     } 
=\wh{D}_N\pa{ \{ \mu^{\e{c}}_{\ell_a} \}_{1}^{N+1} ; \{ \la^{\e{c}}_a \}_{1}^{N} }
\wh{\mc{G}}_N\pa{ \{ \mu^{\e{c}}_{\ell_a} \}_{1}^{N+1} ; \{ \la^{\e{c}}_a \}_{1}^{N}  } \;,
\enq
 into the products of its so-called smooth part 
$\wh{\mc{G}}_N$  and discreet part $\wh{D}_N$. 

\vspace{3mm}
The smooth  part reads:
\beq
\wh{\mc{G}}_N\pa{ \{ \mu^{\e{c}}_{\ell_a} \}_{1}^{N+1} ; \{ \la^{\e{c}}_a \}_{1}^{N}  } = 
\mc{W}_{N}\pab{  \{ \mu^{\e{c}}_{\ell_a} \}_{1}^{N}  }{ \{ \la^{\e{c}}_a \}_{1}^{N} } 
\; \pl{a=1}{N}  \abs{ \f{ \la_a^{\e{c}}-\mu^{\e{c}}_{\ell_{N+1}} -ic }{ \mu^{\e{c}}_{\ell_a}-\mu^{\e{c}}_{\ell_{N+1} } -ic  } }^2 
\cdot  \f{ \ddet{N }{\de_{jk}+U_{jk} }  \ov{ \ddet{N}{\de_{jk}+U_{jk} } } }
{ \ddet{N+1}{\Xi^{\pa{\mu}}} \cdot \ddet{N}{\Xi^{\pa{\la}}} } \;.
\enq
There 
\beq
\mc{W}_{N} \pab{ \paa{z_a}_1^{N} }{ \paa{w_a}_1^{N} }  = 
\pl{a,b=1}{N} \f{ \pa{z_{a}-w_b-ic} \pa{w_{a}-z_{b}-ic}  }{ \pa{z_{a}-z_{b}-ic} \pa{w_{a}-w_{b}-ic}  } \;.
\enq
The discreet part takes the form 
\beq 
\wh{D}_N\pa{ \{ \mu^{\e{c}}_{\ell_a} \}_{1}^{N+1} ; \{ \la^{\e{c}}_a \}_{1}^{N} } = 
\f{ \prod_{k=1}^{N} \paa{4 \sin^{2}\!\big[ \pi F_{ \{\ell_a\} }( \la_k^{\e{c}} )  \big] }  }
{ \pl{a=1}{N+1} \Big\{ 2\pi L \wh{\xi}^{\prime}_{\{\ell_a\}} ( \mu_{\ell_a}^{\e{c}} )  \Big\} 
				\pl{a=1}{N} \Big\{ 2\pi L \wh{\xi}^{\prime}\!\pa{\la_a^{\e{c}} }  \Big\} }
\cdot \pl{a=1}{N} \paf{\mu_{\ell_a}^{\e{c}}-\mu_{\ell_{N+1}}^{\e{c}} }  
{ \la_a^{\e{c}}-\mu_{\ell_{N+1}}^{\e{c}} }^2  \cdot  \det_{N}^2\pac{ \f{1}{\mu_{\ell_a}^{\e{c}}-\la_b^{\e{c}} } }  \;.
\label{definition partie discrete}
\enq
In the remainder of this subsection we discuss the large-$L$ behavior of these two quantities. 

\subsubsection*{The smooth part}

$\wh{\mc{G}}_N$ is called the smooth part as its thermodynamic limit $\mc{G}_n$ only depends on the value of the rapidities
of the particles $\{\mu_{p_a}\}_1^n$ and holes $\{\mu_{h_a}\}_1^n$ entering in the description of the thermodynamic limit of the excited state. We recall that these are defined as in \eqref{definition particle trou thermo lim}. 
The function $\mc{G}_n$ can be readily expressed \cite{KozKitMailSlaTerThermoLimPartHoleFormFactorsForXXZ} 
in terms of  the thermodynamic limit $F$ \eqref{ecriture limite thermo fction shift} of the shift function associated to the excited state
$\{\mu_{\ell_a}^{\e{c}}\}_1^{N+1}$:
\beq
\wh{\mc{G}}_N\pa{ \{ \mu^{\e{c}}_{\ell_a} \}_1^{N+1} ; \{ \la_a^{\e{c}} \}_1^{N} } = \mc{G}_n\pab{ \{ \mu_{p_a} \} }{ \paa{\mu_{h_a}} } \pac{F} 
\times \pa{1+\e{O}\pa{L^{-1}}} \; ,
%
%\quad \e{with} \quad F\pa{\la} \equiv F\pabb{\la}{ \paa{\mu_{p_a}} }{ \paa{\mu_{h_a}} } 
%
%= -Z\pa{\la} - \phi\pa{\la,q} - \sul{a=1}{n} \pac{ \phi\pa{\la,\mu_{p_a}} - \phi\pa{\la,\mu_{p_a}}    }\;, 
%
\enq
with
\bem
\mc{G}_n\pab{ \{ \mu_{p_a} \} }{ \paa{\mu_{h_a}} } \pac{F}= 
\pl{a=1}{n} \pl{\eps=\pm}{} \paa{ \f{   \mu_{h_a}-q+\eps ic }{ \mu_{p_a}-q+\eps ic   } 
\f{ \ex{2i\pi   C\pac{F}\pa{\mu_{h_a} + \eps ic}    }    } {  \ex{2i\pi   C\pac{F}\pa{\mu_{p_a} + \eps ic} }  } }
\cdot \f{  \ex{- 2i\pi  \sul{\eps=\pm}{} C\pac{F}\pa{q + \eps ic}    }  }{   \det^2\pac{I-\tf{K}{2\pi}}  }   \ex{C_0\pac{F}} 
 \\
 \times \mc{W}_n\pab{ \{ \mu_{p_a} \} }{ \paa{\mu_{h_a}} }
% 
% \pl{a,b=1}{n} \f{\pa{\mu_{p_a} - \mu_{h_b}-ic} \pa{\mu_{h_a}-\mu_{p_b}-ic} }{ \pa{\mu_{p_a}-\mu_{p_b}-ic} \pa{\mu_{h_a}-\mu_{h_b}-ic} } \;  
%
\cdot   \ddet{ \msc{C}_{q} }{I+U\pac{F} \big( \{\mu_{p_a}\}_1^n, \{\mu_{h_a}\}_1^n \big)   } 
				\det_{ \msc{C}_{q} }\big[ I+\ov{U}\,\pac{F}\big( \{\mu_{p_a}\}_1^n, \{\mu_{h_a}\}_1^n \big) \big]  \; .
\label{formule explicite G+ thermo}
\end{multline}
There $C\pac{F}$ is the Cauchy transform on $\intff{-q}{q}$ and $C_0\pac{F}$ is given by a double integral
\beq
C\pac{F}\pa{\la} = \Int{-q}{q} \f{\dd \mu}{2i\pi} \f{ F\pa{\mu} }{\mu-\la} \qquad \e{and} \qquad 
C_0\pac{F} = -\Int{-q}{q} \f{ F\pa{\la} F\pa{\mu} }{  \pa{\la-\mu -ic}^2 }   \dd \la \dd \mu  \;.
\label{appendix themo FF definition transfo Cauchy et C0}
\enq
All determinants appearing in \eqref{formule explicite G+ thermo} are Fredholm determinants 
of integral operators of the type $I+A$. The integral operator $I-\tf{K}{2\pi}$ acts on $\intff{-q}{q}$. 
The integral kernels $U$ and $\ov{U}$ are given by 
\beq
U\pa{\om,\om^{\prime}} \pac{F} = \f{-1}{2\pi}\f{\om-q}{\om-q+ic}
\pl{a=1}{n} \Bigg\{ \f{ (\om- \mu_{p_a}) \pa{\om- \mu_{h_a}+ic} }
{\pa{\om- \mu_{h_a}} (\om- \mu_{p_a}+ic) }   \Bigg\}  \cdot \ex{C\pac{2i\pi F}\pa{\om} - C\pac{2i\pi F}\pa{\om + ic}  }
 \f{ K\pa{\om-\om^{\prime}} }{ \ex{-2i\pi F\pa{\om}}-1 }
\label{formule noyau integral U}
\enq
and
\beq
\ov{U}\pa{\om,\om^{\prime}} \pac{F} =\f{1}{2\pi} \f{\om-q}{\om-q - ic}
\pl{a=1}{n} \Bigg\{ \f{ (\om- \mu_{p_a}) \pa{\om- \mu_{h_a}-ic} }
{\pa{\om- \mu_{h_a}} (\om- \mu_{p_a}-ic) }   \Bigg\}  \cdot \ex{C\pac{2i\pi F}\pa{\om} - C\pac{2i\pi F}\pa{\om - ic}  }
 \f{ K\pa{\om-\om^{\prime}} }{ \ex{2i\pi F\pa{\om}}-1 } \;.
\label{formule noyau integral U bar}
\enq
Above, so as to lighten the notations, we have kept the dependence on the particles' and holes' rapidities implicit. 
The operators $I+U\pac{F}$ and $I+\ov{U}\pac{F}$ should de understood as 
acting on function defined on a counterclockwise contour $\msc{C}_{q}$ surrounding the interval $\intff{-q}{q}$
but not any other singularity of the integrand. In particular, the poles at $\om=\mu_{h_a}$ are located inside of $\msc{C}_q$
whereas  the zeroes of $\la\mapsto \ex{-2i\pi F\pa{\la}}-1$
are located outside of the contour. In section \ref{SousSection propriete determinant Fredholm} below we provide a more precise
definition of these determinants, as, in principle, the existence of such a contour is not guaranteed for all possible choices
of parameters $\{\mu_{p_a}\}$, $\{\mu_{h_a}\}$.

\subsubsection*{The discreet part}

The name discreet part originates in that the  leading thermodynamic behavior of $\wh{D}_N$
not only depends on the "macroscopic" rapidities $\{ \mu_{p_a} \}$ and $\paa{\mu_{h_a}}$ entering in the description of the excited state
but also on the set of integers $\paa{p_a}$ and $\paa{h_a}$ characterizing the excited state.
By using the techniques developped in 
\cite{KozKitMailSlaTerThermoLimPartHoleFormFactorsForXXZ,KozKitMailSlaTerEffectiveFormFactorsForXXZ,SlavnovFormFactorsNLSE}
one readily shows that the leading in L thermodynamic behavior of $\wh{D}_N$ takes the form
\beq
\wh{D}_N\pa{ \{ \mu_{\ell_a}^{\e{c}} \}_1^{N+1} ; \{ \la^{\e{c}}_a \}_1^{N} } 
=
D_0\pac{F} \mc{R}_{N,n}\pab{ \{ \mu_{p_a} \}; \paa{p_a} }{ \paa{\mu_{h_a}}; \paa{h_a}  }\pac{F} \times  \pa{1+\e{O}\paf{\ln L}{L}}
\label{AppendixThermolim thermolim D+}
\enq
where
\beq
D_0\!\pac{\nu} =  \f{2q}{2\pi}  \cdot 
\f{\pa{\kappa_-\!\pac{\nu} }^{\nu_-} }{ \pa{\kappa_+\!\pac{\nu} }^{\nu_+ + 2} }
  \pl{a=1}{n} \paf{\la_{N+1}-\mu_{p_a}}{\la_{N+1}-\mu_{h_a}}^2
 \f{ G^2\pa{1-\nu_-}G^2\pa{2+\nu_+}  }{  \pa{2\pi}^{\nu_+ - \nu_-} \cdot \pac{ 2q L \xi^{\prime}_+ }^{\pa{\nu_+ + 1}^2 + \nu_-^2}  }  
\cdot \ex{\f{1}{2} \Int{-q}{q} \f{\nu^{\prime}\!\pa{\la}\nu\!\pa{\mu}-\nu^{\prime}\!\pa{\mu}\nu\!\pa{\la} }{\la-\mu} \dd \la \dd \mu }
 \; ,
\label{AppendixThermoLimD+zero}
\enq
The parameter $\la_{N+1}$ appering above is defined as the unique solution to $L \xi_{F}\pa{\la_{N+1}}= N+1$, $G$ is the Barnes function and
\beq
\kappa\pac{\nu}\pa{\la} = \exp\Bigg\{  -\Int{-q}{q} \f{\nu\pa{\la}-\nu\pa{\mu}}{\la-\mu} \dd \mu \Bigg\} \;.
\label{definition fonction kappa}
\enq
Finally, we agree upon,  
\bem
\mc{R}_{N,n}\pab{ \{ \mu_{p_a} \}; \paa{p_a} }{ \paa{\mu_{h_a}}; \paa{h_a}  }\pac{F}  =
\pl{a=1}{n} \Bigg\{ \f{ \vp\pa{\mu_{h_a},\mu_{h_a}}  \vp\pa{\mu_{p_a},\mu_{p_a}} \ex{\al\pa{\mu_{p_a}}} }
 { \vp\pa{\mu_{p_a},\mu_{h_a}}  \vp\pa{\mu_{h_a},\mu_{p_a}} \ex{\al\pa{\mu_{h_a}}}  }  \Bigg\}
\f{ \pl{a<b}{n} \vp^2\pa{\mu_{p_a},\mu_{p_b}} \vp^2\pa{\mu_{h_a},\mu_{h_b}} }
                                        { \pl{a\not= b }{n} \vp^2\pa{\mu_{p_a},\mu_{h_b}} }
\det^{2}_{n}\pac{ \f{ 1}{ h_a-p_b}}
\\
\hspace{1cm}\times \pl{a=1}{n} \paf{\sin \pac{ \pi \nu\pa{\mu_{h_a}} } }{\pi   }^2  \cdot 
\Ga^{2}\pab{ \{ p_a-N-1+\nu(\mu_{p_a}) \} , \paa{p_a}, \paa{N+2-h_a -\nu\pa{\mu_{h_a}}} , \paa{h_a +\nu\pa{\mu_{h_a}}} }
{ \paa{p_a-N-1} , \{ p_a+\nu(\mu_{p_a}) \}, \paa{N+2-h_a} , \paa{h_a}  } \;.
\label{definition fonctionelle RNn}
\end{multline}
There
\beq
\al\pa{\om} = 2 \nu\pa{\om} \ln \paf{\vp\pa{\om,q} }{ \vp\pa{\om,-q} }
+2 \Int{-q}{q}  \f{\nu\pa{\la}-\nu\pa{\om}}{\la-\om} \dd \la  \; \qquad  \e{and} \quad
\vp\pa{\la,\mu}= 2\pi \f{ \la-\mu }{ p\pa{\la} - p\pa{\mu} }  \; .
\label{definition fonction aleph et varphi}
\enq
Above, we have used the standard hypergeometric-type representation for products of $\Ga$-functions:
\beq
\Ga\pab{ \paa{a_k} }{  \paa{ b_k } }= \pl{k=1}{n} \f{ \Ga\pa{a_k} }{ \Ga\pa{b_k} } \; .
\enq

\subsection{The Fredholm determinants}
\label{SousSection propriete determinant Fredholm}

In this section we provide a way to define Fredholm determinants entering in the leading asymptotic behavior of the 
properly normalized form factors of the conjugated field in the case where the contour $\msc{C}_q$, as it has been described previously, does not exist. 
Acutally, this definition holds as well in the case of compex valued rapidities. 
Prior to stating the result, we need to introduce some notations. 
Given $\de>0$ and $\eps>0$, we introduce 
\beq
U_{\de} = \Big\{ z \in \Cx \; : \; \abs{\Im\pa{z}} < \de \Big\} \qquad  \e{and} \qquad 
\msc{K}_{\eps}= \Big\{ z \in \Cx  \; : \;  \abs{\Im \pa{z} } < \tf{\de}{2} \; \e{and} \; \abs{\Re \pa{z} } < q+\eps \Big\} \;.
\enq
Finally, given $\be_0 \in \Cx$, we denote
\beq
\bs{U}_{\be_0} = \paa{  z \in \Cx  \; : \;  10 \Re\pa{\be_0} \geq \Re\pa{z}\geq \Re\pa{\be_0} \;\; \e{and} \;\; \abs{\Im\pa{z}} \leq \Im\pa{\be_0}  }
\enq
and agree that $D_{0,\eps}$ stands for the open disk of radius $\eps$ that is centered at $0$. 
Also, $\ov{S}$ refers to the closure of the set $S$. 

\begin{prop}
\label{Proposition Holomorphie en beta et rapidite part-trou det fred}

Let $m\in \mathbb{N}$ be fixed and $\eps, \de>0$ be small enough. Assume that one is given 
two  holomorphic function $\nu$ and $h$  on $U_{2\de}$,  such that 
\beq
h\pa{ U_{2\de} }\subset \paa{ z \; : \; \Re\pa{z} >0 } \quad \e{and}  \quad z \mapsto \Im \pa{h\pa{z}} \; \e{is} \; \e{bounded} \;  \e{on} \; U_{2\de} \,.
\enq
\noindent Then, there exists
\begin{itemize}

\item $\be_0 \in \Cx$ with $\Re\pa{\be_0}>0$ large enough  and $\Im\pa{\be_0}>0$ small enough
\item $\ga_0 > 0$ but small enough
\item  a small counterclockwise loop $\msc{C}_q$ around $\ov{\msc{K}}_{\eps}$ and in $U_{2\de}$
\end{itemize}
such that given $\nu_{\be}\pa{\la}=\nu\pa{\la}+ i \be h\pa{\la} $, one has 
\beq
 \ex{-2i\pi \ga \pa{\nu+ i\be h}\pa{\la}}-1 \not= 0  \qquad \forall \la \quad \e{on}\; \e{and} \;\e{inside} \; \msc{C}_{q} 
 \quad \e{and} \quad \e{uniformly} \; \e{in}
\pa{ \be , \ga}  \in \bs{U}_{\be_0} \times D_{0,\ga_0} \; .
\enq
Moreover, given an integral kernel $U\big[  \ga \nu_{\be} \big] \big(  \{ \mu_{p_a} \}_1^n,\paa{\mu_{h_a}}_1^n \big)(\om,\om^{\prime})$ 
as defined by \eqref{formule noyau integral U}, the function
\beq
\mc{F}\big( \bs{z} \big)  = 
G\big(1- \ga \nu_{\be}\pa{-q} \big) G\big(2+  \ga \nu_{\be} \pa{q}\big)  \pl{a=1}{n} \pa{ \ex{-2i\pi \ga \nu_{\be}\pa{\mu_{h_a}}} - 1 } \; \cdot \;
\ddet{ \msc{C}_q }{ I+ \ga U\big[  \ga \nu_{\be} \big] \big(  \{ \mu_{p_a} \}_1^n,\paa{\mu_{h_a}}_1^n \big)  } \;
\label{appendix thermo lim FF fonction avec limite themor reg}
\enq
is holomorphic in $\bs{z} = \big( \{ \mu_{p_a} \}_1^n,\paa{\mu_{h_a}}_1^n,\be, \ga \big)$
belonging to $\mc{D}_0=U_{\de}^{n} \times \msc{K}_{\eps}^n \times   \bs{\wt{U}}_{\be_0} \times D_{0,\ga_0}$, 
this uniformly in $0\leq n\leq m$.

It admits a (unique) analytic continuation to $\mc{D} = U_{\de}^{n} \times \msc{K}_{\eps}^n \times 
\paa{ z  \in \Cx \; : \; \Re\pa{z} \geq -\eps} \times D_{0,1+\eps} $.

\end{prop}

\Proof

We begining by proving the first statement. 
We choose a small counterclockwise loop $\msc{C}_q$ around $\ov{\msc{K}}_{\eps}$ and in $U_{2\de}$. We denote by 
$K$ the compact such that $\Dp{}K=\msc{C}_q$. Then one has, $\forall \la \in K$ 
\beq
\Im\big(\nu_{\be}\pa{\la} \big) \geq - \sup_{K} \abs{\Im\pa{\nu\pa{\la}}} - \Im\pa{\be_0} \sup_{K} \abs{\Im\pa{h(\la)}} +
\Re\pa{\be_0} \inf_K \pac{ \Re\pa{h(\la)} }
\enq
Thus, $\Im\big(\nu_{\be}\pa{\la}\big) >0$ prodided that $\be \in \bs{U}_{\be_0}$, with $\Im\pa{\be_0}=\de $ and 
$\Re\pa{\be_0}$ such that
\beq
\Re\pa{\be_0} >  \f{ 1 }{ \inf_K \pac{ \Re\pa{h(\la)} }   }
  \Big[ \sup_{K} \abs{\Im\pa{\nu\pa{\la}}} + \Im\pa{\be_0} \sup_{K} \abs{\Im\pa{h(\la)}}    \Big] \;.
\enq

Then, $\ga_0$ is chosen such that 
\beq
0< \ga_0 \leq \f{1}{2} \pac{  \sup_K \abs{\nu\pa{\la} } + \pa{ 10 \Re\pa{\be_0} + \Im\pa{\be_0} } \sup_K\abs{h\pa{\la}}  }\; .
\enq
It is then easy to show that, for such a $\ga_0$,  one has $\ga_0 \sup_K\abs{\nu_{\be}(\la)}\leq \tf{1}{2}$. 
This estimate holds uniformly in $\pa{\be,\ga}\in \bs{U}_{\be_0}\times D_{0,\ga_0}$. As a consequence,  
the function $\la \mapsto \vp\pa{\la,\be,\ga}$ with 
\beq
\vp\pa{\la,\be,\ga} =  \ex{-2i\pi \ga \nu_{\be}\pa{\la}}-1 \; ,
\enq
has no zeroes in $K$.

Hence, the integral kenrel of the operator  $\ga U\big[\ga \nu_{\be}\big] \big(  \{ \mu_{p_a} \}_1^n,\paa{\mu_{h_a}}_1^n \big)$ is 
smooth on $\msc{C}_q\times \msc{C}_q$. As $\msc{C}_q$ is compact, this aforementioned operator is trace class 
on $L^1( \msc{C}_q)$. Moreover, it depends holomorphically on $\{\mu_{p_a}\}_1^n\in U_{\de}^n$,  
$\{\mu_{h_a}\}_1^n\in \msc{K}_{\eps}^n$ and $\pa{\be,\ga}\in \bs{U}_{\be_0}\times D_{0,\ga_0}$. 
Standard properties of operator detereminants \cite{SimonsInfiniteDimensionalDeterminants} then ensure that $\mc{F}(\bs{z})$,
as defined in \eqref{appendix thermo lim FF fonction avec limite themor reg} is 
holomorphic in $\bs{z} \in \mc{D}_0$. 
We remind that, for the purpose of this section, a bold letter $\bs{z}$  refers to
vectors of the type $\bs{z} = \big(  \{\mu_{p_a}\}_1^n, \{\mu_{h_a}\}_1^n, \be , \ga \big)$.

Let $A$ be the set 
\beq
A= \bigg\{ \bs{z} \in \mc{D} \; : \; \pl{\eps=\pm}{} \ga^{-1} \pa{ \ex{-2i\pi \ga \nu_{\be}(\eps q)}    -1 } 
\pl{a=1}{n}   \ga^{-1} \pa{ \ex{-2i\pi \ga \nu_{\be}(\mu_{h_a})}    -1 }  = 0  \bigg\} \;. 
\enq
By definition $A$ is an analytic set. Moreover since it is realized as the locus of zeroes of a single, non-zero, holomorphic function on $\mc{D}$, 
it has at least codimension 1, \textit{cf} \cite{ScheidemannIntroSeveralComplexVars}. 

It follows from the first part of the proof that $\mc{F}\big( \bs{z} \big)$ is indeed well defined on $\mc{D}_0$. It can be 
naturally extended to a holomorphic function on the set $\mc{D}\setminus A$ by deforming the original contour 
$\msc{C}_q$ in such a way that the zeroes of $\ex{-2i\pi \ga \nu_{\be}\pa{\la}}-1$ are not surrounded by $\msc{C}_q$ whereas
the points $\pm q$ and $\mu_{h_1}, \dots, \mu_{h_n}$ are surrounded by it. Such a deformation is always possible as, on the one 
hand, $\bs{z}\notin A$ so that the zeroes of  $\ex{-2i\pi \ga \nu_{\be}\pa{\la}}-1$ are indeed distinct from 
the points $\pm q$ and $\mu_{h_1}, \dots, \mu_{h_n}$. On the other hand, it is allowed to deform the contour by applying the Cauchy theorem:
the integrand is a holomorphic function on the region where the deformation of interest takes place. Indeed,
the only terms that are not explicitly holomorphic in the integral kernel \eqref{formule noyau integral U} 
are the  various Cauchy transforms. However, as $\nu_{\be}$ is holomorphic on $U_{2\de}$, the cut of the Cauchy 
transform can be deformed within $U_{2\de}$ as long as it keeps its endpoints on $\pm q$. 

It thus remains to show that $\mc{F}(\bs{z})$ can be analytically continued through $A$. For this, it is enought to 
show that given any $\bs{z}^{\pa{0}} \in A$, there exists an open neighborhood $U$ of $\bs{z}^{\pa{0}}$ such that 
setting $W = \pa{\mc{D}\setminus A}\cap U$, $\mc{F}_{\mid W}$ is bounded.

We parameterize $\bs{z}^{\pa{0}} \in A$ as $\bs{z}^{\pa{0}} = 
\pa{  \{\mu_{p_a}^{\pa{0}}\}_1^n, \{\mu_{h_a}^{\pa{0}}\}_1^n, \be^{\pa{0}} , \ga^{\pa{0}}}$.  
This means that if $\bs{\rho} \in \mc{D}\setminus A$ and is sufficiently close to $\bs{z}^{\pa{0}}$, 
there exists zeroes (not necessarily distinct) $z_1\pa{\bs{\rho}},\dots, z_{\ell}\pa{\bs{\rho}}$
of $\la  \mapsto \vp\big( \la,\be,\ga \big) $
that will approach $\pm q$ or $\mu_{h_1}^{\pa{0}}, \dots, \mu_{h_n}^{\pa{0}}$ in the limit
 $\bs{\rho} \tend \bs{z}^{\pa{0}}$ in $\mc{D}\setminus A$. 
 
Indeed, the zeroes  of a holomorphic function form discreet sets. Hence, there exists a contour 
$\Ga_{z^{\pa{0}}}$ consisting of small counterclockiwse circles around  $\pm q $ and  $\mu_{h_1}^{\pa{0}}, \dots, \mu_{h_n}^{\pa{0}}$
such that $ \vp\big(\la,\be^{\pa{0}},\ga^{\pa{0}} \big) \not= 0  $ for all $\la \in \Ga_{z^{\pa{0}}}$. 

The function $ \pa{\la,\be,\ga} \mapsto \vp\pa{\la, \be,\ga}$ is continuous and $\Ga_{z^{\pa{0}}}$ is compact. 
Hence, there exists an open neighborhood $B^{\pa{0}}$ of 
$\big(\be^{\pa{0}} , \ga^{\pa{0}}\big)$ in $\Cx^2$, such that $\vp\pa{\la,\be,\ga} \not=0$ for any $\la \in \Ga_{z^{\pa{0}}}$ and 
$\pa{\be,\ga} \in B^{\pa{0}}$. As a consequence, we get that 
for any $\pa{\be,\ga} \in B^{\pa{0}}$, the number of zeroes (counted with their multiplicities) of $\la \mapsto \vp\pa{\la,\be,\ga}$ 
is constant and equal to some integer $\ell$.

Let $V_0$ be an open set contained in the bounded connected component of $\Cx \setminus \Ga_{z^{\pa{0}}}$ and let 
$\bs{\rho} = \pa{  \{\mu_{p_a}\}_1^n, \{\mu_{h_a}\}_1^n, \be , \ga} \in \mc{D}\setminus A$  be such that $\mu_{h_a}\in V_0$ for any $a=1,\dots,n$
and $\pa{\be,\ga} \in B^{\pa{0}}$. As $\bs{\rho} \in \mc{D}\setminus A$, we necessarily have that the zeroes of $\la\mapsto \vp\pa{\la,\be,\ga}$ all differ from $\pm q$ and $\mu_{h_a}$, $a=1,\dots,n$. 
By deforming, if necessary, the initially introduced contour $\msc{C}_q$, we can represent the Fredholm determinant by its Fredholm series:
\beq
\ddet{\msc{C}_q}{ I+ \ga U\big[\ga \nu_{\be} \big] ( \{\mu_{p_a}\}_1^n ; \{\mu_{h_a}\}_1^n )  } = \sul{m \geq 0}{} \f{1}{m!} \Int{\msc{C}_q}{} \dd^n \om 
\ddet{m}{ U_{\be} \pa{\om_a,\om_b}}
\pl{a=1}{m} \f{ \ga }{ \ex{-2i\pi \ga \nu_{\be}\pa{\om_a}} -1 } \;,
\enq
where we have set 
\beq
U_{\be} \pa{\om,\om^{\prime}}= 
\f{- 1}{2\pi}\f{\om-q}{\om-q+ic}
\pl{a=1}{n} \paa{ \f{ (\om- \mu_{p_a}) (\om- \mu_{h_a}+ic) }
{(\om- \mu_{h_a}) (\om- \mu_{p_a}+ic) }   }  \cdot 
\ex{C[ 2i\pi \ga \nu_{\be} ]\pa{\om} - C[ 2i\pi \ga \nu_{\be} ]\pa{\om + ic}  }
 K\pa{\om-\om^{\prime} } \;. 
\enq
We set  $\wt{\msc{C}} = \msc{C}_q \cup \Ga_{ \bs{z}^{\pa{0}}  } $. Due to the symmetry of the integrand, 
we may carry out the substitution
\beq
\f{1}{m!}\Int{ \wt{\msc{C}} \cup \{ -\Ga_{ \bs{z}^{\pa{0}} \} }  }{ } \dd^m \om  = \sul{s=0}{m} \f{1}{s!\pa{m-s}!} 
\Int{ \wt{\msc{C} } }{} \dd^{m-s} \om \Int{ \{ -\Ga_{ \bs{z}^{\pa{0}} \} } }{} \pl{j=1}{s} \dd \om_{m-j+1}
\enq
Note that  $-\Ga_{ \bs{z}^{\pa{0}}  }$ appearing above stands  for the contour $\Ga_{ \bs{z}^{\pa{0}}  }$ but endowed with the opposite orientation. 
Further, notice that for any  symmetric function $f\pa{\om_1,\dots,\om_s}$ that is holomorphic in a neighborhood of the points $z_j(\bs{\rho})$and vanishing on the diagonals ($\om_{\ell}= \om_{p}$ for $\ell\not=p)$,  one has
\beq
\Int{ -\Ga_{ \bs{z}^{\pa{0}}  } }{} f\pa{\om_1,\dots,\om_s} \pl{a=1}{s} \f{ \ga   }{ \ex{-2i\pi\ga  \nu_{\be}\pa{\om_s}}-1  }  \; \cdot  \dd^s \om = 
s! \sul{ \substack{ \intn{1}{\ell} = \a_-\cup \a_+ \\  \abs{\a_+}=s } }{} \hspace{-4mm} 
f\pa{z_{\a_1}(\bs{\rho}),\dots, z_{\a_s}(\bs{\rho}) } 
\pl{j=1}{s} \f{1}{ \nu^{\prime}_{\be}\big( z_{\a_j}(\bs{\rho}) \big)} \; .
\label{calcul pole ac fction sym}
\enq
Above, the sum runs through all the  partitions of $\intn{1}{\ell}$ into two disjoint subsets $\a_+\cup \a_-$ such that 
$\a_+=\pa{\a_1,\dots,\a_s}$ contains $s$ elements, \textit{ie} $\abs{\a_+}=s$. 
Note that we have here tacitly assumed that all of the roots are simple. The case of multiple roots can then be obtained by carrying out 
a limiting procedure on \eqref{calcul pole ac fction sym}. 

Therefore, we obtain the below representation for the Fredholm determinant
\bem
\ddet{\msc{C}_q}{ I+ \ga U\big[ \ga \nu_{\be} \big]  } = \sul{m \geq 0}{}  \sul{s=0}{ \min\pa{m,p}} \f{ \pa{-1}^m }{\pa{m-s}!} 
 \sul{ \substack{ \intn{1}{p} = \a_-\cup \a_+ \\  \abs{\a_+}=s } }{} \Int{ \wt{\msc{C}}}{} \f{ \dd^{m-s} \om }{\pa{2\pi}^{m}}
  \ddet{m}{ K(\om_k-\om_j) }\\
\pl{b=1}{s}\paa{  \f{z_{\a_b}(\bs{\rho})-q}{ z_{\a_b}(\bs{\rho})-q+ic} 
\pl{a=1}{n} \pac{ \f{ ( z_{\a_b}(\bs{\rho}) - \mu_{p_a} ) \pa{ z_{\a_b}(\bs{\rho}) - \mu_{h_a}+ic} }
				{\pa{ z_{\a_b}(\bs{\rho}) - \mu_{h_a}} ( z_{\a_b}(\bs{\rho}) - \mu_{p_a}+ic )}   }
\f{ \ex{C\pac{ 2i\pi \ga \nu_{\be} }\pa{ z_{\a_b}(\bs{\rho}) } - C\pac{ 2i\pi \ga \nu_{\be} }\pa{z_{\a_b}(\bs{\rho}) + ic}  }}
{  \nu^{\prime}_{\be}\pa{ z_{\a_b}(\bs{\rho}) } } }  \\
\pl{k=1}{m-s}\paa{  \f{\om_k-q}{\om_k-q+ic} 
\pl{a=1}{n} \pac{ \f{ (\om_k- \mu_{p_a}) \pa{\om_k- \mu_{h_a}+ic} }{\pa{\om_k- \mu_{h_a}} (\om_k- \mu_{p_a}+ic} )   }
\ga \f{ \ex{C\pac{ 2i\pi \ga \nu_{\be} }\pa{\om_k} - C\pac{ 2i\pi \ga \nu_{\be} }\pa{\om_k + ic}  }}{ \ex{-2i\pi \nu_{\be}\pa{\om_k}} -1 } }  
\label{calcul determinant Fredholm I+U}
\end{multline}
and we agree upon the shorthand notation $\om_{m-j+1}=z_{\a_j}(\bs{\rho})$ for $j=1,\dots,s$ for the 
determinant that occurs in the first line.

For any fixed $\bs{\rho}$, one has the decompotition in respect to zeroes
\beq
\ex{-2i\pi \ga \nu_{\be}\pa{\om}}-1 = \pl{a=1}{ \ell } \pa{\om-z_a(\bs{\rho})} \cdot V_{\be,\ga}\pa{\om} \; ,
\enq
with $V_{\be,\ga}\pa{\om}$ a holomorphic function on $\msc{K}_{\eps}$ that has no zeroes on $\wt{\msc{C}}$  and $V_0$
uniformly in  $\pa{\be,\ga}\in B^{\pa{0}}$. It thus follows that the function 
\beq
\pl{a=1}{n} \paa{ \pa{ \ex{-2i\pi \ga \nu_{\be}(\mu_{h_a})}-1 }  \pl{b=1}{s}  \f{ 1 }{ z_{\a_b}(\bs{\rho})-\mu_{h_a} } }
\enq
is bounded on $\bs{\rho}$ belonging to  $ \pa{\mc{D}\setminus A} \cap \msc{U}$ with $\msc{U} = U_{\de}^n \times V_0^n \times B^{\pa{0}}$. 
Note that the above reasoning holds for simple roots. In the case of multiple roots, one should first carry out a limiting 
procedure on the level of \eqref{calcul determinant Fredholm I+U}, which will lead to the appearence of derivatives. 
The final conclusion however still holds. We leave these details to the reader.

It only remains to focus on the Cauchy transforms. The latter can be represented as
\beq
\ex{C\pac{2i\pi \ga \nu_{\be}}\pa{\om}} = \exp\bigg\{  \Int{-q}{q}   \ga \f{ \nu_{\be}\pa{\la}- \nu_{\be}\pa{\om}  }{ \la-\om} \dd \la  \bigg\} 
\paf{\om-q}{ \om+q}^{ \ga \nu_\be\pa{\om}} \;.
\enq
As a consequence, the only divergencies that can arize from the Cauchy transform are located at $\om= \pm q$. 

If there exists a $k$ such that $z_k(\bs{z}^{\pa{0}})=\pm q$, then there exists $\ell_k \in \mathbb{Z}$ such that 
$\ga \nu_{\be}( z_k(\bs{z}^{\pa{0}}) )= \ell_k$. As a consequence, the Cauchy transforms occuring in the second line of 
\eqref{calcul determinant Fredholm I+U} may introduce divergent contributions. 
Yet, since the Barnes' function has a simple zero of order $p+1$ at $-p$, with $p \in \mathbb{N}$, it is easy to see that  
\beq
G\pa{1-\ga \nu_{\be}\pa{-q}} G\pa{2+\ga \nu_{\be}\pa{q}} 
 \pa{z_k(\bs{\rho})  -q} \cdot \pl{k=1}{\ell} \paf{ z_k(\bs{\rho}) -q}{ z_k(\bs{\rho}) +q}^{\ga \nu_\be\pa{z_k(\bs{\rho})}} 
\enq
is bounded for $\bs{\rho} \in  \pa{\mc{D}\setminus A}\cap \msc{U}$. 
The fact that all other terms in \eqref{calcul determinant Fredholm I+U} are bounded is evident. 
The theorem then follows after applying the analytic continuation theorem in many variables \cite{ScheidemannIntroSeveralComplexVars}. 

\qed

\section*{Conclusion}

In this paper, we proved the convergence towards naturally associated quantities in the continuous model 
of scalar products and form factors arizing in the lattice discretization 
of the NLSM. This provides the last missing step 
towards the proof of determinant-based representations for these object in the continuum. Our approach was 
based on a generalization and simplification of the techniques proposed in \cite{DorlasOrthogonalityAndCompletenessNLSE}. 
We have also provided a unambiguous procedure for defining the class of Fredholm determinants that occurs in the 
large volume limit of properly normalized form factors  in integrable models, this on the example 
of the NLSM. 
It would be quite natural to continue this kind of considerations for lattice discretizations of more
involved models such as the Sine-Gordon model. However, in this case additional complications will arize due to 
the non-conservations of the number of particles.

\section*{Acknowledgment}

I acknowledge the support of the EU Marie-Curie Excellence Grant MEXT-CT-2006-042695.
I would like to thank N.Kitanine, J.-M. Maillet, N. Slavnov, J. Teschner and V. Terras for stimulating discussions.

%%%%%%%%%%%%%%%%%%%%%%%%%%%%%%%%%%%%%%%%%%%%%%%%%%%%%%%%%%%%%%%%%%%%%%%%%%%%%%%%%%%%%%%%%%%%%%%%%%%%%%%%%%%%%%%%%%%%%%%%%%%%%%%%%%%%%%%%%%%%%%%%%%%%%%%%%
%%%%%%%%%%%%%%%%%%%%%%%%%%%%%%%%%%%%%%%%%%%%%%%%%%%%%%%%%%%%%%%%%%%%%%%%%%%%%%%%%%%%%%%%%%%%%%%%%%%%%%%%%%%%%%%%%%%%%%%%%%%%%%%%%%%%%%%%%%%%%%%%%%%%%%%%%
%%%%%%%%%%%%%%%%%%%%%%%%%%%%%%%%%%%%%%%%%%%%%%%%%%%%%%%%%%%%%%%%%%%%%%%%%%%%%%%%%%%%%%%%%%%%%%%%%%%%%%%%%%%%%%%%%%%%%%%%%%%%%%%%%%%%%%%%%%%%%%%%%%%%%%%%%

\appendix

\section{Proof of theorem}
\label{Appendix Proof continuous limit}

\subsection{Combinatorial representation of the eigenstates}

\begin{lemme}
\label{Lemme decomposition combinatoire etat de Bethe}
Let $\{\la_j\}$ be $N$ generic parameters, then the below representation holds:
\beq
B\pa{\la_1}\dots B\pa{\la_N}\ket{  0 } = \sul{ \substack{ 1 \leq n_1 \leq  \\  \dots \leq n_N \leq M}  }{}
\pl{a=1}{M} \f{1}{ \pa{\# \Ga_a}!} \cdot f_{\{ \la \}} \pa{n_1,\dots, n_N} \cdot  \be^{*}_{n_1} \dots \be^*_{n_N} \ket{0}
\label{lemme combinatoire pour vecteur propre}
\enq
where we agree upon $\be_k^{*} = -i \sqrt{c} \chi^*_k \rho_{Z_k}$  and $\Ga_k =\paa{ \ell \; : \; n_{\ell}= k}$, $k=1,\dots,M$. 
In \eqref{lemme combinatoire pour vecteur propre}, we have set 
\bem
f_{\{ \la \}}\pa{n_1,\dots, n_N } = \sul{ \sg \in \mf{S}_N }{} 
\pl{ a<b }{N} \f{ \la_{\sg\pa{a}  } - \la_{\sg\pa{b}} + i c  \e{sgn}\pa{n_b-n_a}  }{ \la_{\sg\pa{a}} - \la_{\sg\pa{b}}  }
\pl{ a=1 }{ N }\paa{ \a\pa{\la_{\sg\pa{a} }} }^{ \pac{ \f{n_a-1}{2} } } \, \paa{ \ov{\a}\pa{\la_{\sg\pa{a} }} }^{ \pac{ \f{M-n_a}{2} } }  \\
\times \pl{ a=1 }{ N }  \paa{ 1- \pa{-1}^{n_a}\De \pa{ \f{c}{4} -i \f{ \la_{\sg\pa{a}} }{ 2 }  } } \;.
\label{definition fonction f lambda des entiers}
\end{multline}
The sign function appearing above has been defined in \eqref{definition fonction signe} and $\pac{\cdot}$ stands for the floor function 
and we agree upon 
\beq
\a\pa{\la} =  \pa{1-\f{c\De}{4} + i \f{ \la \De}{4} }  \pa{1+\f{c\De}{4} + i \f{ \la \De}{4} } \qquad \e{and} \qquad 
\ov{\a}\pa{\la} =  \a\pa{-\la} \;.
\enq

\end{lemme}

\Proof 
It is a standard fact \cite{KorepinNormBetheStates6-Vertex} that, for any generic set of parameters $\{\la_j\}$, 
the action of a product of  B operators 
on the pseudo-vaccum can be expressed as a sum over all the possible partitions of the set $\intn{1}{N}$ into $M$ non-intersecting sets 
$\Ga_{1}, \dots , \Ga_{M}$:  

\beq
\pl{k=1}{N} B\pa{\la_k}\ket{  0 } = \sul{ \substack{ \intn{1}{N} \\ = \cup_{k=1}^{M} \Ga_k } }{}
\pl{\ell=2}{ M } \pl{m=1}{\ell-1} \Bigg\{ 
\pl{a \in \Ga_m}{}  \pl{b \in \Ga_{\ell} }{}   \f{\la_{a} - \la_{b} + ic }{ \la_a-\la_b }  
\pl{a \in \Ga_{\ell} }{} \bigg( Z_m + i \f{\De \la_a}{2}   \bigg)
\pl{b \in \Ga_{m} }{} \bigg( Z_{\ell} - i \f{\De \la_b }{2}  \bigg)  \Bigg\}  \cdot 
\pl{a=1}{M} \pa{ \be^*_a}^{\# \Ga_a} \ket{0} \;.
\label{ecriture decomposition combinatoire vecteur Bethe}
\enq
We stress that in the above decomposition, the ordering of the partition counts, \textit{ie} $\paa{1,2,3}\cup \paa{\emptyset}$
is different from $\paa{\emptyset} \cup \paa{1,2,3}$. Also, we have denoted by $\# \Ga_a$ the cardinality of the set $\Ga_a$.

Note that there is a one-to-one correspondence between the set of all such partitions and choices of $N$ integers $n_1,\dots, n_N$
in $\intn{1}{M}$ by the formula $\Ga_k = \paa{ \ell \; : \; n_{\ell}= k }$. 
One can thus recast the sums in \eqref{ecriture decomposition combinatoire vecteur Bethe} as ones over such choices of integers. 
Namely,
\beq
\pl{k=1}{N} B\pa{\la_k}\ket{  0 } = \sul{ \substack{ 1\leq n_1\leq  \\ \dots \leq n_N\leq M } }{ }
\pl{a=1}{ M} \f{1}{ \pa{\# \Ga_a}! } \cdot f_{\{ \la \}} \pa{n_1,\dots n_N}
\cdot  \be^*_{n_1} \dots \be^*_{n_M} \ket{0} \;
\label{ecriture decomposition combinatoire vecteur Bethe}
\enq
where 
\beq
f_{\{ \la \}} \pa{n_1,\dots n_N} = \sul{\sg\in \mf{S}_N }{} 
\pl{\ell=2}{ M } \pl{m=1}{\ell-1} 
\Bigg\{  \pl{ \substack{ a  : \\ n_{\sg^{-1}\pa{a}}=m }  }{}  \pl{\substack{ b  : \\ n_{\sg^{-1}\pa{b}}=\ell } }{}  
\paa{  \f{\la_{a} - \la_{b} + ic }{ \la_a-\la_b }  }
\pl{ \substack{ a  : \\ n_{\sg^{-1}\pa{a}}=\ell }  }{}\hspace{-3mm} \bigg( Z_m + i \f{\De \la_a}{2}   \bigg)
\pl{ \substack{ b  : \\ n_{\sg^{-1}\pa{b}}=m }  }{} \hspace{-3mm} \bigg( Z_{\ell} - i \f{\De \la_b }{2}   \bigg)
 \Bigg\}  \;.
\label{ecritute f lambda discret comme somme permutations}
\enq
In \eqref{ecriture decomposition combinatoire vecteur Bethe}, we have recast the sum over $n_a \in \intn{1}{M}$ into one over 
the ordered choices of integers $1\leq n_1\leq \dots \leq n_N\leq M$, this
by introducing an additional sum over permutations $\sg\in \mf{S}_N$ in \eqref{ecritute f lambda discret comme somme permutations}. 
However, so as not to count elements twice, for each given 
choice of integers $1\leq n_1\leq \dots \leq n_N\leq M$, we divide by $\prod_{a=1}^{N} \pa{\# \Ga_a}!$. Indeed, the permutation group leaves
the diagonals unaltered, and each of such diagonals corresponds to elements in the set $\Ga_a$, occuring in the partition $\cup \Ga_a$. 

Then, it is enough to observe that 
\beq
\pl{\ell=2}{M} \pl{m=1}{\ell-1} 
\pl{ \substack{ a  : \\ n_{\sg^{-1}\pa{a}}=m }  }{}  \pl{\substack{ b  : \\ n_{\sg^{-1}\pa{b}}=\ell } }{}  
\paa{ \f{\la_{a} - \la_{b} + ic }{ \la_a-\la_b }   }
= \pl{a<b}{N} \f{ \la_{\sg\pa{a}}-\la_{\sg\pa{b}} + ic \e{sgn}\pa{n_b-n_a}  }{ \la_{\sg\pa{a}}-\la_{\sg\pa{b}}  }
\enq
and 
\beq
\pl{\ell=2}{M} \pl{m=1}{\ell-1}  
\pl{ \substack{ a  : \\ n_{\sg^{-1}\pa{a}}=\ell }  }{}\hspace{-3mm} \bigg( Z_m + i \f{\De \la_a}{2}   \bigg)
\pl{ \substack{ b  : \\ n_{\sg^{-1}\pa{b}}=m }  }{} \hspace{-3mm} \bigg( Z_{\ell} - i \f{\De \la_b }{2}   \bigg)
= \pl{ a=1 }{ N }\paa{ \a\pa{\la_{\sg\pa{a} }} }^{ \pac{ \f{n_a-1}{2} } } \, \paa{ \ov{\a}\pa{\la_{\sg\pa{a} }} }^{ \pac{ \f{M-n_a}{2} } }  
\times \pl{ a=1 }{ N }  \paa{ 1- \pa{-1}^{n_a}\De \pa{ \f{c}{4} -i \f{ \la_{\sg\pa{a}} }{ 2 }  } }
\nonumber 
\enq

\qed

In fact, given a solution $\{ \la_{\ell_a}  \}_1^N $ of the Bethe equations \eqref{ecriture log BAE}, the associated function
$f_{\{ \la \}}$ as defined in \eqref{definition fonction f lambda des entiers} is bounded uniformly in $\De$ small enough. This 
is an important property in respect to taking the $\De\tend 0$ limit.

\begin{lemme}
\label{Lemme estimation borne fonction f lambda}
Let $\{ \la_{\ell_a}  \}_1^N $ be a solution  of the Bethe equations \eqref{ecriture log BAE} associated with the choice of integers 
$\ell_1< \dots< \ell_N$. Then, there exists $\De_0$ small enough and a constant $C_{ \{ \ell_a \} }$ solely depending on $N, L$, $\De_0$
and the choice of integers $\{ \ell_a \}$, such that 
\beq
\abs{ f_{\{ \la \}}\pa{n_1,\dots, n_N} } \leq C_{\{\ell_a\}} \qquad  \e{uniformly} \quad \De \in \intff{0}{\De_0} \; , 
\enq
where $f_{\{\la\}}$ has been defied in \eqref{definition fonction f lambda des entiers} \;.
\end{lemme}

\Proof

It follows from the continuity in $\De$ on $\intff{0}{\De_0}$ of $\De \mapsto \la_{\ell_a}$, $a=1,\dots,N$, 
(\textit{cf} subsection \ref{SousSectionSpectrumEigenvectors} ) that the function 
$\De \mapsto \min_{a<b}\abs{\la_{\ell_a}-\la_{\ell_b}}$ is continuous on $\intff{0}{\De_0}$. Thus, it attains its minimum
at some $\wt{\De} \in \intff{0}{\De_0}$. However, in virtue of the repulsion principle \eqref{repulsion principle}, 
this minimum must be strictly positive, and thus 
\beq
m_{\{\ell_a\}} = \inf_{\De \in \intff{0}{\De_0}} \min_{a<b} \abs{\la_{\ell_a}-\la_{\ell_b}} >0 \;.
\enq
For each $\De$, the associated parameters $\la_{\ell_a}$ are bounded. Hence,  the function 
$\De \mapsto \max_a \abs{\la_{\ell_a}}$ is well defined and continuous in $\De \in \intff{0}{\De_0}$. As 
argued before, this implies that
\beq
M_{\{  \ell_a \}}  = \sup_{\De \in \intff{0}{\De_0}}  \max_a \abs{\la_{\ell_a}} < +\infty \;.
\enq

Hence, given any choice of integers $1\leq n_1\leq \dots \leq n_N\leq M$,  
\beq
\abs{ \pl{  n_a }{ N }  \paa{ 1- \pa{-1}^{n_a}\De \pa{ \f{c}{4} -i \f{ \la_{\sg\pa{a}} }{ 2 }  } } } \leq
\pa{1 +  \f{\De}{4} \pa{c+2 M_{\{\ell_a\}}} }^N
\enq
and, for any $a \in \intn{1}{n}$
\beq
\abs{ \a\pa{ \pm  \la_{a }} } \leq \ex{ M_{\{\ell_a\}} \De + \f{c \De}{2} } \;.
\enq
Thus, as $M\De = L$
\beq
\abs{ \pl{ a=1 }{ N }\paa{ \a\pa{\la_{\sg\pa{a} }} }^{ \pac{ \f{n_a-1}{2} } } \, 
\paa{ \ov{\a}\pa{\la_{\sg\pa{a} }} }^{ \pac{ \f{M-n_a}{2} } } } 
\leq 
\ex{ \pa{  M_{\{\ell_a\}}  + \f{c }{2} } \De \sul{a=1}{N} \pac{ \f{n_a-1}{2} }  +   \pac{ \f{M-n_a}{2} }  }
\leq \ex{ \pa{  M_{\{\ell_a\}}  + \f{c }{2} } \f{N L}{2}  } \;.
\enq
Last but not least
\beq
\abs{ \pl{ a<b }{N} \f{ \la_{\sg\pa{a}  } - \la_{\sg\pa{b}} + i c  \e{sgn}\pa{n_b-n_a}  }{ \la_{\sg\pa{a}} - \la_{\sg\pa{b}}  } } 
\leq  \paf{ 2M_{ \{\ell_a\} } + c }{ m_{ \{\ell_a\} } }^{\f{N\pa{N-1}}{2} } \;.
\enq
Putting all these  estimates together leads to:
\beq
\abs{ f_{\{\la \}} \pa{n_1,\dots,n_N} } \leq  N! \pa{1 +  \f{\De_0}{4} \pa{c+2 M_{\{\ell_a\}}} }^N\ex{ \pa{  M_{\{\ell_a\}}  + \f{c }{2} } \f{N L}{2}  }     \paf{ 2M_{ \{\ell_a\} } + c }{ m_{ \{\ell_a\} } }^{\f{N\pa{N-1}}{2} }   \; ,
\enq
uniformly in $\De \in \intff{0}{\De_0}$ and $1 \leq n_1 \leq \dots \leq n_N \leq M$.  \qed 

\subsection{The scalar product formula}
\label{appendix Scalar products proof}

By setting $\be^*_k=-i \sqrt{c} \chi^*_k \rho_{Z_k}$, \, $\be_k=i \sqrt{c} \rho_{Z_k} \chi_k $ and 
$h_k=Z_k+ \chi^*_k\, \chi_k$ it is easy to see that these satisfy
\beq
\pac{ \be_k, \be^*_n} = \De c \, h_k \, \de_{kn} \qquad \e{and} \qquad  \pac{ h_k, \be^*_n} = \f{\De c}{2} \be^*_k \de_{kn} \;.
\enq
These commutation relations readily lead to 
\beq
\bra{0} \big( \be_k \big)^n \cdot \big( \be_{p}^{*} \big)^m  \ket{ 0 } =  \de_{n,m} \pa{ \de_{k,p}  + \de_{n,0}\de_{m,0}} \pa{\De c}^n  \, n! \, 
\pl{\ell=1}{n} \pa{ Z_k + \pa{\ell-1}\tf{\De c }{4} }. 
\enq

Thus, given a solution of the Bethe equations  $\{\la_{\ell_a} \}_1^N$ defined by the integers $\ell_1<\dots< \ell_N$, 
and a set of generic parameters $\{\mu_a\}_1^N$ that are bounded, and satisfy the condition $\min_{a<b}\abs{\mu_a-\mu_b}>0$, 
building on the representation for the Bethe vectors  \eqref{lemme combinatoire pour vecteur propre}, one gets
\beq
\braket{ \psi\pa{  \{\mu_a \}_1^N } }{ \psi\pa{  \{\la_{\ell_a} \}_1^N } } = \pa{\De c }^N 
\sul{ \substack{ 1 \leq n_1  \leq \\ \dots \leq n_N\leq M } }{}
\ov{f_{\{\mu\}}}\pa{n_1,\dots,n_N}  f_{\{\la\}}\pa{n_1,\dots,n_N}  
\pl{a=1}{M} \pl{p=1}{\# \Ga_a-1}\f{ Z_{n_a} + \tf{p\De c}{4} }{ p+1 } \;.
\label{equation calcul PS etat Bethe combinatoire}
\enq
Note that, to obtain \eqref{equation calcul PS etat Bethe combinatoire}, we have used that 
\beq
\bra{0} \be_{n_1^{\prime}}  \dots  \be_{n_N^{\prime}}   \be_{n_1}^{*}  \dots  \be_{n_N}^{*}  \ket{0} \not= 0 \qquad \e{with}
\quad 
1\leq n_1^{\prime} \leq \dots n_N^{\prime} \leq M \quad \e{and} \quad 1\leq n_1 \leq \dots n_N \leq M\; ,
\enq
only if $n_a=n_a^{\prime}$ for any $a$. 
It is convenient to split in \eqref{equation calcul PS etat Bethe combinatoire} 
the contributions form the diagonals ($n_a=n_{a+1}$ for some $a$) from those lying purely off the diagonal:
\beq
\braket{ \psi\pa{  \{\mu_a \}_1^N } }{ \psi\pa{  \{\la_{\ell_a} \}_1^N } } = L_1 + L_2 \; ,
\enq
where 
\beq
L_1 = \pa{\De c }^N  \sul{ \substack{ 1 < n_1  < \\ \dots < n_N\leq M } }{}
\ov{f_{\{\mu\}}}\pa{n_1,\dots,n_N}  f_{\{\la\}}\pa{n_1,\dots,n_N}  
\pl{a=1}{N} Z_{n_a}  \;, 
\enq
and 
\beq
L_2 = \pa{\De c }^N 
\sul{k=1}{N-1} \sul{ \substack{ 1 \leq n_1  \leq \dots \\ \leq n_k=n_{k+1}\leq \\ \dots  \leq n_N\leq M } }{} 
\ov{f_{\{\mu\}}}\pa{n_1,\dots,n_N}  f_{\{\la\}}\pa{n_1,\dots,n_N}  
\pl{a=1}{N} \pl{p=1}{\# \Ga_a-1}\f{ Z_{n_a} + \tf{p\De c}{4} }{ p+1 } \;.
\enq
The multiplicative factor in $L_2$ is bounded due to 
\beq
\abs{ \pl{a=1}{N} \pl{p=1}{\# \Ga_a-1}\f{ Z_{n_a} + \tf{p\De c}{4} }{ p+1 }  } \leq 
\pl{a=1}{N} \pl{p=1}{\# \Ga_a} \f{1+ \tf{\De c p}{ 4 } }{  p  } \leq \ex{N \De M \f{c}{4}} \;.
\enq
By applying lemma \ref{Lemme estimation borne fonction f lambda} to the function 
$f_{\{\la\}}\pa{n_1,\dots,n_N}$  and carrying out a similar reasoning to the one of the lemma, we get 
$\ov{f_{\{\mu\}}}\pa{n_1,\dots,n_N} \leq C_N\pa{\{\mu\}}$, uniformly in $\De \in \intff{0}{\De_0}$ and for some $\mu_a$
dependent constant  $C_N\pa{\{\mu\}}$. Hence, as $L=\De M$
\beq
\abs{L_2} \leq N  \pa{\De c }^N  M^{N-1} C_{\{\ell_a\}} \cdot C_N\pa{\{\mu\}} \ex{\f{NL}{4}c} =  \e{O}\pa{\De} \; . 
\enq
As a consequence, $L_2$ does not contribute to the $\De\tend 0$ limit of the scalar product.

It remains to treat $L_1$. Using that the parameters $\la_{\ell_a}$ are all continuously differentiable in respect to $\De$, 
it is readily seen that, uniformly in the choices $1 < n_1 < \dots < n_N \leq M$ and 
$y_k \in \intof{ x_{n_k-1} }{ x_{n_k} }$ with $x_{p}=p\De$, one has
\beq
f_{\{ \la \}} \pa{n_1,\dots,n_N} =  c^{-\f{N}{2}} \ex{\f{i\pi}{2} N } 
\vp\big( y_1,\dots, y_N \mid \{ \la^{\e{c}}_{\ell_a} \}_1^N \big) \cdot  \pa{1+\e{O}\pa{\De}}
\enq
and likewise
\beq
f_{\{ \mu \}} \pa{n_1,\dots,n_N} = c^{-\f{N}{2}} \ex{\f{i\pi}{2} N } \vp\pa{y_1,\dots, y_N \mid \{ \mu_{a} \}_1^N} \cdot \pa{1+\e{O}\pa{\De}} \;.
\enq
As a consequence, 
\beq
L_1 = \Int{0}{L}\ov{ \vp\pa{y_1,\dots, y_N \mid \{ \mu_{a} \}} }  \cdot 
\vp\big( y_1,\dots, y_N \mid \{ \la^{\e{c}}_{\ell_a} \} \big) \cdot  g_{\De}\pa{y_1,\dots, y_N}  \;  \dd^N \! y \; ,
\label{equation representation integrale discrete du PS}
\enq
with 
\beq
g_{\De}\pa{y_1,\dots, y_N}  = \sul{ \substack{ 1\leq n_1 <  \\  \dots < n_N \leq M } }{} 
\pl{k=1}{N} \bs{1}_{ \intof{x_{n_k-1} }{ x_{n_k} } }\pa{y_k}  \; \cdot \, \pa{1+\e{O}\pa{\De}} \;.
\enq
where $\bs{1}_{\intof{a}{b}}\pa{x}$ denotes the indicator function of the interval $\intof{a}{b}$. 

It is readily seen that, for $\De$ small enough,  $\sup_{\intff{0}{L}^N}\abs{ g_{\De} } \leq 2$, 
that $g_{\De} \in L^{1}\big( \intff{0}{L}^N \big)$ and that, almost everywhere
\beq
g_{\De}\pa{y_1,\dots, y_N} \tend \bs{1}_{  \mc{D} } \pa{y_1,\dots, y_N } \qquad \e{where} \qquad 
\mc{D} = \Big\{ \pa{y_1,\dots, y_N}  \; : \;   0\leq y_1 < \dots < y_N \leq L \Big\} \;.
\enq
As both functions $\vp$ are bounded on $\intff{0}{L}^N$, we are in  position to apply the dominated convergence theorem:
\eqref{equation representation integrale discrete du PS} converges to the \textit{rhs} of 
\eqref{equation convergence PS discret vers integrale}.

We have thus proven that the scalar product defined in terms of products of B operators and their adjoints does converge,
in the $\De \tend 0$ limit, to the scalar product of the continuous model. However, as follows from 
theorem \ref{Theorem Nikita Scalar Products}, such scalar products admits a finite-size $N$ determinant representation. 
It is straightforward to compute the $\De \tend 0$ limit of the $rhs$ in \eqref{ecriture produits scalaires} hence obtaining
the determinant representation for the scalar products in the continuous model.  \qed

\subsection{The form factors of the conjugated field operator}

In order to prove theorem \ref{Theorem cvgce lattice discreization}, we first notcie that 
the restrictions of the operators $\chi_k$ and $\chi^*_k$ to the $N$-particle Hilbert space 
$\msc{H}_N = \e{Vect}\paa{ \chi^*_{n_1} \dots \chi_{n_N}^* \ket{0} \; , \; 1 \leq n_1 \leq \dots \leq n_N \leq M }$
are bounded operators:
\beq
\chi_k : \msc{H}_N \tend  \msc{H}_{N-1}  \quad \norm{\chi_k}_{N,N-1} = \sqrt{N\De} 
\qquad \e{and} \qquad 
\chi_k : \msc{H}_N \tend  \msc{H}_{N+1}  \quad \norm{\chi_k^*}_{N,N+1} = \sqrt{\pa{N+1}\De} \;.
\enq
Above, $\norm{\cdot}_{N,N+1}$ stands for the operator norm on linear operators from $\msc{H}_N$ to $\msc{H}_{N+1}$. 
It then follows that, for $\De$ small enough, 
\beq
\norm{ \tau^{-1}\!\!\pa{\nu} \cdot B\pa{\nu} - \tf{\be_M^{*}}{2} }_{N,N+1} = \e{O}\big( \De^{\tf{3}{2}} \big) \;.
\label{ecriture estimation norme operateur}
\enq
There, $\tau^{-1}\!\!\pa{\nu} \cdot B\pa{\nu}$ is given by \eqref{ecriture reconstruction Oota} and $\be_k^*$ is as defined in lemma \ref{Lemme decomposition combinatoire etat de Bethe}. 

The bound \eqref{ecriture estimation norme operateur} follows from the fact that all the operators $\rho_{Z_k}$
are bounded on $\msc{H}_{N}$ and that they can be represented there, for $\De$ small enough, are uniformly
convergent series. The rest follows from standart estimates of bounded operator-values series.

One can then represent the form factor as
\beq
\mc{F}^{\pa{\De}}_{ \{\la\} ; \{\mu\} } = \bra{ \psi\pa{  \{\mu_{\ell_a} \}_1^{N+1} } } \tau^{-1}\!\pa{\nu} B\pa{\nu} \ket{ \psi\pa{  \{\la_{r_a} \}_1^N } } 
= \mc{F}^{\pa{\De,1}}_{ \{\la\} ; \{\mu\} }  \; + \;   \mc{F}^{\pa{\De,2}}_{ \{\la\} ; \{\mu\} } \;, 
\enq
where 
\beq
\mc{F}^{\pa{\De,2}}_{ \{\la\} ; \{\mu\} }  = 
\bra{ \psi\pa{  \{\mu_{\ell_a} \}_1^{N+1} } } \paa{ \tau^{-1}\!\pa{\nu} B\pa{\nu} - \tf{ \be^*_M}{2}  }\ket{ \psi\pa{  \{\la_{r_a} \}_1^N } } 
\enq
and 
\beq
\mc{F}^{\pa{\De,1}}_{ \{\la\} ; \{\mu\} }  = \f{1}{2}
\bra{ \psi\pa{  \{\mu_{\ell_a} \}_1^{N+1} } }  \be^*_M  \ket{ \psi\pa{  \{\la_{r_a} \}_1^N } }  \;.
\enq
The Cauchy-Schwarz formula leads to 
\beq
\abs{ \mc{F}^{\pa{\De,2}}_{ \{\la\} ; \{\mu\} } } \leq \norm{ \psi\pa{ \{ \mu_{\ell_a}\}_1^{N+1} } } \cdot  
\norm{ \psi\pa{ \{ \la_{r_a}\}_1^{N} } } \cdot \norm{ \tau^{-1}\!\pa{\nu} B\pa{\nu} - \tf{\be_M^{*}}{2} }_{N,N+1}  
= \e{O}\big( \De^{\f{3}{2}} \big) \;.
\enq
There we have used the results following from section \ref{appendix Scalar products proof} that norms of Bethe vectors are bounded 
uniformly in $\De$ small enough and the estimates \eqref{ecriture estimation norme operateur}. 

It remains to analyse the limit of $\abs{ \mc{F}^{\pa{\De,1}}_{ \{\la\} ; \{\mu\} } }$. By computing the scalar products likewise 
to what has been done in the previous section, we obtain 
\beq
 \mc{F}^{\pa{\De,1}}_{ \{\la\} ; \{\mu\} }  = \f{ \pa{\De c }^{N+1}  }{ 2 } 
\sul{ \substack{ 1 \leq n_1  \leq \\ \dots \leq n_N\leq M } }{}
\ov{f_{\{\mu\}}}\pa{n_1,\dots,n_N,M}  f_{\{\la\}}\pa{n_1,\dots,n_N}  
\pl{a=1}{M} \pl{p=1}{\# \Ga_a-1}\f{ Z_{n_a} + \tf{p\De c}{4} }{ p+1 } \;.
\enq
Above, the sets $\Ga_a$ are subordinate to the sequence of inegers $\paa{ n_1,\dots, n_N,M=n_{N+1} }$. 
Very similar estimates and calculations to those gathered in sub-section \ref{appendix Scalar products proof},
lead to the conclusion that 
\beq
\De^{-1} \cdot \mc{F}^{\pa{\De,1}}_{ \{\la\} ; \{\mu\} }  \limit{ \De }{ 0 }
-\f{i \sqrt{c}}{2} \Int{0}{L} 
\ov{ \vp\big(y_1,\dots, y_N, L \mid \{ \mu^{\e{c}}_{\ell_a} \} \big) } \cdot 
\vp\big(y_1,\dots, y_N \mid \{ \la^{\e{c}}_{r_a} \}\big) \bs{1}_{  \mc{D} } \pa{y_1,\dots, y_N }  \; \dd^N \! y \;.
\enq

Since, $\De^{-1} \mc{F}^{\pa{\De,2}}_{ \{\la\} ; \{\mu\} } \tend 0$ in the $\De \tend 0$ limit, we get that indeed, 
$\De^{-1}  \mc{F}^{\pa{\De}}_{ \{\la\} ; \{\mu\} }$ does indeed converge to the form factor of the operator
$-i \sqrt{c} \tf{ \Phi^{\dagger}\pa{0} }{2}$ in the continuous model. 
The determinant representation for the form factor of the $\Phi^{\dagger}$ operator in the continous model
then follows from taking the $\De \tend 0$ limit on the determinant representation given in proposition \ref{Proposition facteur de forme Psi}, which is 
straightforward.

\end{document}